\documentclass{article}
\usepackage{PRIMEarxiv}
\usepackage{alias}
\usepackage[round]{natbib}
  
\title{Variational Bayes Portfolio Construction}

\author{
  Nicolas Nguyen\thanks{Corresponding author: \texttt{nicolas.nguyen@uni-tuebingen.de}} \\
  University of Tübingen\thanks{This work was initiated during an internship at Capital Fund Management (CFM).} \\
   \And
  James Ridgway \\
  Capital Fund Management (CFM) \\
  \And
  Claire Vernade \\
  University of Tübingen \\
}

\begin{document}
\maketitle

\begin{abstract}
Portfolio construction is the science of balancing reward and risk; it is at the core of modern finance. In this paper, we tackle the question of optimal decision-making within a Bayesian paradigm, starting from a decision-theoretic formulation. Despite the inherent intractability of the optimal decision in any interesting scenarios, we manage to rewrite it as a saddle-point problem. Leveraging the literature on variational Bayes (VB), we propose a relaxation of the original problem. This novel methodology results in an efficient algorithm that not only performs well but is also provably convergent. Furthermore, we provide theoretical results on the statistical consistency of the resulting decision with the optimal Bayesian decision. Using real data, our proposal significantly enhances the speed and scalability of portfolio selection problems. We benchmark our results against state-of-the-art algorithms, as well as a Monte Carlo algorithm targeting the optimal decision.
\end{abstract}

\bibliographystyle{plainnat}

\section{Introduction}
\label{sec:introduction}
Portfolio construction (or selection) is a fundamental problem in modern finance \citep{Markovitz1952,merton1972analytic}, involving the strategic allocation of capital across multiple assets to achieve an optimal tradeoff between risk and return. As financial markets grow in complexity, designing robust portfolios that effectively account for market uncertainty has become increasingly critical. Traditional approaches, such as Markowitz’s mean-variance optimization \citep{Markovitz1952}, have provided a foundational framework for portfolio construction but are now facing challenges in modern finance problems. Markets are becoming increasingly dynamic, with non-Gaussian asset returns, and, in some cases, small dataset sizes. This has led to suboptimal performance in real-world scenarios (see discussions in \citet{benichou2016agnostic}). Formally, the mean-variance framework of a portfolio of $d$ assets can be stated as choosing weights $\delta$ from a decision set\footnote{For example, we might consider the $d$-dimensional simplex as a decision set, $\cD = \Delta_d := \{\delta = (\delta_i)_{i\in[d]}\in\real^d :\,\delta_i\geq 0\quad\forall i\in[d]\,,\,\sum_{i=1}^d \delta_i = 1\}$.} $\cD\subset \real^d$ by solving
\begin{align}\label{eq:Markovitz}
&\argmax_{\delta \in \mathcal{D}} \delta^T\mu&&\text{s.t.}\quad \delta^T \Sigma\delta \leq \lambda\,,
\end{align}
where $\mu\in\real^d$ is the mean of observations, $\Sigma\in\real^{d\times d}$ its covariance matrix, and $\lambda$ is a risk tolerance parameter. 
Extensive research has focused on improving this framework by addressing key challenges, such as incorporating higher-order moments of return distributions \citep{harvey2010portfolio}, introducing robust optimization techniques \citep{ismail2019robust}, and refining covariance matrix estimation from noisy data \citep{benichou2016agnostic,bun2017cleaning,agrawal2022covariance,benaych2023optimal}. The Markowitz portfolio provides a systematic approach to balancing return and risk, and despite its limitations, continues to serve as a hard-to-beat benchmark.

Beyond variance-focused methods, utility-based portfolio construction considers an investor’s subjective perception of risk and reward, allowing for a nuanced approach to decision-making that can address concerns such as tail risks or other features not captured by variance alone. A particularly useful utility function in this context is the exponential utility function with risk parameter $\lambda > 0$, defined as
\begin{align}\label{eq:exponential_utility}
    u_\lambda : (y, \delta) \mapsto \frac{1}{\lambda}(1 - e^{-\lambda y^\top \delta})\,,
\end{align}
which has the following remarkable property: when returns are Normally distributed, maximizing the expected exponential utility is equivalent to the mean-variance optimization in \eqref{eq:Markovitz} \citep{merton1969lifetime},
\begin{align*}
\argmax_{\delta \in\cD} \E{y\sim\cN(\mu, \Sigma)}{u_\lambda(y, \delta)}\;\Longleftrightarrow  \;\argmax_{\delta\in\cD}\big\{\mu^\top\delta - \lambda \delta^\top\Sigma\delta\big\}\,.
\end{align*}
This equivalence establishes a strong link between the mean-variance theory and utility-based approaches, making the latter a compelling alternative for capturing investor preferences \citep{gerber1998utility}. However, this equivalence does not hold beyond Normally distributed returns, since the expected utility function cannot be computed in closed-form. Despite recent advances on this problem \citep{luxenberg2024portfolio}, the general problem of \emph{decision in the face of uncertainty} remains. Specifically, beyond point estimates, we need a reliable estimator of the mean and covariance $(\mu,\Sigma)$ of the recorded historical time series. 

\textbf{Contributions.} We introduce a novel approach to portfolio construction, departing from traditional utility-based methods by adopting a Bayesian framework \citep{barry1974portfolio,black1992global,de2019bayesian,kato2024general}. Specifically, we reformulate this task as a versatile \emph{minmax optimization problem}, which can be efficiently addressed using Variational Bayes (VB)  (\textbf{\cref{sec:method_principle}}), while providing formal consistency guarantees (\textbf{\cref{sec:theoretical_guarantees}}). Furthermore, we instantiate our proposed algorithm ($\algVB$) for several relevant models (\textbf{\cref{sec:algorithm}}) that we test in practice on real data (\textbf{\cref{sec:experiments}}), showing that it achieves state-of-the-art performance in several settings. 

\textbf{Notations.} For an arbitrary probability space $(\Theta, \mathcal{T}_\Theta, \pi)$ where $\Theta$ is a Polish space, we denote as $\cMplus(\Theta)$ the set of probability distributions on $\Theta$. Throughout this paper, $\pi$ denotes a generic probability distribution, where its probability space will be clear from context. $\cS^d$ is the set of squared positive definite matrices of size $d$. For a probability distribution $\pi$ and a random variable $\theta$, $\E{\rho}{\theta}$ denotes the expectation of $\theta$ when $\theta\sim\rho$. The Kullback-Leibler (KL) divergence between two probability distributions $\pi_1$ and $\pi_2$ is denoted $\KL(\pi_1, \pi_2)$. For a random variable $\theta$ and a measure $\pi$, we use the infinitesimal notation $\pi(\dint\theta)$, generalizing the notation $\pi(\theta = \cdot)$ when $\Theta$ is countable.
\section{Problem Setting}
\label{sec:problem_setting}
We consider a supervised learning setting where we have access to $n$ observations $H_n = (Y_t)_{t\in[n]}$, where each $Y_t\in\real^d$. We assume that $(Y_t)_t$ is the realization of a stochastic process parameterized by an unknown parameter $\theta^*$, $(Y_t)_{t}\sim P_{\theta^*}$. Until \cref{sec:algorithm}, we do not make any additional assumption on $P_{\theta^*}$ for now. We define a probability space $(\Theta, \cT_\Theta)$ associated with $\theta$ and express our initial uncertainty about this parameter through a prior distribution $\pi_0$. 

Building on the discussions in \cref{sec:introduction}, we formalize our portfolio construction problem in the lens of Bayesian decision theory \citep[Chapter 2]{robert2007bayesian}. The \emph{Bayesian decision} $\delta^*$ with respect to a utility function $u$ is the decision $\delta\in\cD$ that maximises the \emph{posterior expected utility} (or \emph{Bayesian risk}),
\begin{align}
\label{eq:optimal_Bayesian_decision}
    \delta^* = \argmax_{\delta \in \cD} \int_{\real^d} u(\Ypred, \delta)\pi(\dint\Ypred\condi H_n)\,,
\end{align}
where $\pi(\dint \Ypred \condi H_n)$ is the \emph{posterior predictive distribution} of new (unseen) observation $\Ypred$. Note that $\delta^*$ is a function of $H_n$, $\delta^*=\delta^*(H_n)$, but this dependency is omitted to simplify notation.
Under the particular choice of exponential utility \eqref{eq:exponential_utility}, we can rewrite \eqref{eq:optimal_Bayesian_decision} as
\begin{align}
\label{eq:optimal_Bayesian_decision_exp}
    \delta^* = \argmin_{\delta \in \cD} \int_{\real^d} e^{-\lambda \delta^\top\Ypred}\pi(\dint\Ypred\condi H_n)\,,
\end{align}
for a given $\lambda > 0$ fixed by the user.
One major challenge in Bayesian modelling is the lack of closed-form solutions for posterior predictive distributions, except for simple statistical models (see \cref{subsec:app_Gaussian_Gaussian}). Hence, directly computing \eqref{eq:optimal_Bayesian_decision_exp} in closed-form is generally infeasible. While various methods exist to numerically compute this integral (\emph{e.g.} Monte-Carlo estimates), they tend to be computationally expensive, particularly in high-dimensional spaces. We address this in the following section by rewriting the objective function as a \emph{saddle point}. We then make use of the same relaxation as in Variational Bayes \citep[VB;][]{jordan1999introduction} to approximate the inner optimisation.
\section{Exponential Utility Maximization as a Saddle-Point Optimization}
\label{sec:method_principle}
\subsection{Main Observation}
Our main contribution is to show that maximizing an exponential utility function is \emph{equivalent} to solving a saddle-point optimization problem. We believe that the following result may be of independent interest to anyone seeking to maximize an exponential utility for various applications.
\begin{theorem}\label{th:main_th}
The optimal Bayesian decision \eqref{eq:optimal_Bayesian_decision_exp} can be written as a saddle-point,
\begin{align}\label{eq:final_min_max}
    \delta^* = \argmin_{\delta \in \cD} \max_{\rho \in \cMplus(\real^d \times \Theta)}\left\{-\KL(\rho, \Tilde{\pi}_n) + Z_\delta\right\}\,,
\end{align}
where $\pi_n$ is the posterior distribution over the joint parameter $(y, \theta)\in\real^d\times\Theta$, $\Tilde{\pi}_n$ is defined as $\dint\Tilde{\pi}_n = \frac{e^{-\lambda \delta^\top \Ypred}}{\E{\pi_n}{e^{-\lambda \delta^\top \Ypred}}}\dint\pi_n$ and $Z_\delta = -\E{\pi_n}{e^{-\lambda \delta^\top \Ypred}}$ is a term that does not depend on $\rho$.
\end{theorem}
The proof of this result relies on a well-known change-of-measure lemma, included below for completeness (see \citet[Lemma 2.2;][]{alquier2024user} for a proof).
\begin{lemma}[Change of measure lemma \citep{donsker1983asymptotic}]
\label{lemma:donsker_varadhan}
For any probability $\pi$ on a probability space $(\mathcal{X},\mathcal{T})$ and any measurable function
  $h : \mathcal{X} \rightarrow \mathbb{R}$ such that $\int{\rm e}^h  \rm{d}\pi < +\infty$,
  \begin{equation*}
    \log\int {\rm e}^{h(x)} \pi(\mathrm{d}x) = \underset{\rho \in \cMplus(\cX)}{\sup}
    \left[ \int h(x) \rho(\mathrm{d}x) -\KL(\rho,\pi) \right]\,.
  \end{equation*}
\end{lemma}
We now prove our main result, which applies this change-of-measure on the log of the exponential utility.
\begin{proof}[Proof of \cref{th:main_th}]
By rewriting \eqref{eq:optimal_Bayesian_decision_exp},
\begin{align*}
\delta^* = \argmin_{\delta \in \cD} \int_{\real^d} e^{-\lambda \delta^\top \Ypred}\pi(\dint\Ypred \condi H_n) &\stackrel{\textcolor{blue}{\rm{(i)}}}{=} \argmin_{\delta \in \cD} \log\int_{\real^d} e^{-\lambda \delta^\top \Ypred}\pi(\dint\Ypred \condi H_n)\\
&\stackrel{\textcolor{blue}{\rm{(ii)}}}{=} \argmin_{\delta \in \cD} \log\int_{\real^d \times \Theta} e^{-\lambda \delta^\top \Ypred}\underbrace{\pi(\dint (\Ypred, \theta) \condi H_n)}_{\textcolor{blue}{:=}\pi_n((\dint (\Ypred, \theta))}\,,
\end{align*}
where in \textcolor{blue}{(i)} we took the log in front of the objective function, and in \textcolor{blue}{(ii)} we marginalized out $\theta$ conditionally on $H_n$ (since $\int_\Theta \pi(\dint(\Ypred, ,\theta)\condi H_n) =\pi(\dint\Ypred\condi H_n)$). 
Applying \cref{lemma:donsker_varadhan} with $h:x \mapsto-\lambda \delta^\top x$ gives
\begin{align}\label{eq:main_trick}
    \delta^* = \argmin_{\delta \in \cD} \sup_{\rho \in \cMplus(\real^d \times \Theta)}\left\{-\lambda \delta^\top \E{\rho}{\Ypred} - \KL(\rho, \pi_n)\right\}\,.
\end{align}
We next have to show that the expression inside the supremum can be expressed as a KL divergence (up to an additive constant that does not depend on $\rho$); we observe that for any $\delta\in\cD$,
\begin{align*}
    -\lambda\delta^\top \E{\rho}{\Ypred} = -\int_{\real^d\times\Theta}\log \frac{1}{e^{-\lambda\delta^\top\Ypred}}\rho(\dint(\Ypred, \theta))
\end{align*}
and therefore, by introducing the probability measure $\Tilde{\pi}_n$ defined as $\dint\Tilde{\pi}_n = \frac{e^{-\lambda \delta^\top \Ypred}}{\E{\pi_n}{e^{-\lambda \delta^\top \Ypred}}}\dint\pi_n$, we have
\begin{align}
\label{eq:transformation_of_inside_sup}
    &-\lambda \delta^\top \E{\rho}{\Ypred} - \KL(\rho, \pi_n) = - \int_{\real^d \times \Theta} \log\left(\frac{\dint \rho}{\dint \Tilde{\pi}_n}(\Ypred, \theta)\right)\rho(\dint (\Ypred, \theta)) + Z_\delta= -\KL(\rho, \Tilde{\pi}_n) + Z_\delta\,.
\end{align}
Combining \eqref{eq:main_trick} with \eqref{eq:transformation_of_inside_sup}, we can rewrite the Bayes optimal decision as
\begin{align*}
    \delta^* = \argmin_{\delta \in \cD} \sup_{\rho \in \cMplus(\real^d \times \Theta)}\left\{-\KL(\rho, \Tilde{\pi}_n) + Z_\delta\right\}\,,
\end{align*}
where the supremum is indeed achieved for $\rho = \Tilde{\pi}_n$. 
\end{proof}
We introduce the \emph{risk} function $\mathcal{R}_{\cMplus}$ over all probability measures;
\begin{align*}
    \forall \delta\in\cD,\quad \mathcal{R}_{\cMplus}(\delta) = \sup_{\rho \in \cMplus(\real^d \times \Theta)}\left\{-\KL(\rho, \Tilde{\pi}_n) + Z_\delta\right\}\,,
\end{align*}
and $\delta^*$ is the decision that minimizes the risk $\mathcal{R}_{\cMplus}$.
\subsection{Variational Bayes Approximation of \texorpdfstring{$\delta^*$}{Lg}}
Computing $\Tilde{\pi}_n$ is challenging because the normalization constant $\E{\pi_n}{e^{-\lambda \delta^\top \Ypred}}$ is intractable for non-conjugate models (for which it is equivalent to computing the integral \eqref{eq:optimal_Bayesian_decision_exp}).
We now demonstrate how the min-max formulation in \eqref{eq:final_min_max} can be leveraged to enable the use of VB approximation.

\textbf{Maximization over a subspace of measures.} Fix an arbitrary decision $\delta \in \cD$. Since $Z_\delta$ does not depend on $\rho$, the distribution that solves the maximum writes\footnote{The negative sign is omitted for now but we will plug it in the final objective function.}
\begin{align}
\label{eq:real_rho}
\rho^*(\delta) := \argmin_{\rho \in \cMplus(\real^d \times \Theta)} \KL(\rho, \Tilde{\pi}_n) \,,
\end{align}
where we emphasize that $\rho^*$ depends on $\delta$ since $\Tilde{\pi}_n$ does.
VB approximations instead solve a restriction of \eqref{eq:real_rho}: we define a family of measures $\cF \subseteq \cMplus(\real^d \times \Theta)$ for which the restricted problem \eqref{eq:real_rho} over this family is considered tractable. For example, the \emph{mean-field family} \citep{parisi1988statistical,bishop2006pattern} assumes independence between parameters: assuming $\Theta$ factorizes as a product of $K\geq 1$ subspaces, $\Theta = \prod_{i=1}^K \Theta_i$, the mean-field family of $(\real^d, \Theta)$ is defined as
\begin{align*}
    &\Fmf\left(\real^d\times \Theta\right) = \left\{ \rho \in \cMplus(\real^d \times \Theta):\,\rho(\dint(\Ypred, \theta)) = \rho_y(\dint \Ypred) \prod_{i=1}^K \rho_{i}(\dint \theta_i) \,:\,\rho_y \in \cMplus(\real^d),\, \rho_i \in \cMplus(\Theta_i)\,\forall i \in [K]\right\}\,.
\end{align*}
Notice that $\Fmf$ does not make \emph{any assumption} on the form of the distributions $\rho_y$ or $(\rho_i)_i$'s, but only relies on the factorisation assumption and the underlying statistical model. We denote by $\rhoVB$ the Mean-field variational approximation of $\rho^*$, that is,
\begin{align}\label{eq:def_rho_vb}
    \rhoVB(\delta) = \argmin_{\rho \in \Fmf(\real^d \times \Theta)} \KL(\rho, \Tilde{\pi}_n) \,.
\end{align}
Since we deal with a parametric underlying statistical model, $\rhoVB$ is also parametric. The main advantage of using $\Fmf$ is that $\rhoVB$ can be computed numerically since it is the solution of a fixed-point equation.
\begin{proposition}\label{prop:fixed_point}
The variational distribution is written as $\rhoVB = \rho_y\otimes_{j=1}^K \rho_j$, where for any $\delta\in\cD$ we have
\begin{align*}
    &\log\rho_y(\dint\Ypred)
    \propto \E{\rho_1,\dots\rho_K}{\log e^{-\lambda \delta^\top \Ypred}\pi(\Ypred, \theta, H_n)} \\
     &\log\rho_j(\dint\theta_j) \propto \E{\rho_y, \bm{\rho}_{-j}}{\log e^{-\lambda \delta^\top \Ypred}\pi(\Ypred, \theta, H_n)}\,,
\end{align*}   
where for any $j\in[K]$, $\E{\bm{\rho}_{-j}}{\cdot}$ denotes the expectation with respect to the measures $(\rho_i)_{i\in[K]\setminus\{j\}}$.
\end{proposition}
The proof of the previous equation is provided in \cref{sec:app_complete_statements}, and is a direct consequence of a well-known result for Mean-field variational inference \citep[Chapter 10;][]{bishop2006pattern}.

\textbf{Minimization over decisions.}
Once we found the variational approximation $\rhoVB$ for a given $\delta$, we define the corresponding \emph{variational decision} $\deltaVB$ by just plugging $\rhoVB$ into \eqref{eq:final_min_max},
\begin{align}\label{eq:def_delta_vb}
    \deltaVB = \argmin_{\delta \in \cD} \obj(\delta)\,,
\end{align}
where we introduced the objective function $\delta\mapsto \obj(\delta)$,
\begin{align}\label{eq:objective_function}
\obj(\delta) &= \sup_{\rho \in \Fmf(\real^d \times \Theta)}\left\{-\KL(\rho, \Tilde{\pi}_n) + Z_\delta\right\} = -\KL(\rhoVB(\delta), \Tilde{\pi}_n) + Z_\delta\,.
\end{align}
Note that $\obj$ can be seen as an approximation of the risk function $\mathcal{R}_{\cMplus}$, where for all $\delta\in\cD$, $\obj(\delta)\leq \mathcal{R}_{\cMplus}(\delta)$. Then, one key observation is that once we computed $\rhoVB$, we don't have to compute $Z_\delta$ because
\begin{align*}
    -\KL(\rhoVB(\delta), \Tilde{\pi}_n) &+ Z_\delta = -\E{\rhoVB}{\log(\rhoVB)} - \lambda\delta^\top \E{\rhoVB}{\Ypred} + \E{\rhoVB}{\log\pi_n} + C\,,
\end{align*}
where $C$ does not depend on $\delta$, and hence \eqref{eq:objective_function} can be computed in closed-form. Since $\rhoVB$ depends on $\delta$, optimizing with respect to $\delta$ requires to alternate Gradient-descent steps on \eqref{eq:objective_function} with adjustment steps \eqref{eq:def_rho_vb} in the following way:

\textbf{i) Gradient-Descent step.} We perform one step of gradient descent with a constant step-size $\eta$,
\begin{align*}
  \hat\delta^{\rm{(k+1)}} \gets \hat\delta^{\rm{(k)}} - \eta \nabla_{\delta} \obj(\deltaVB^{\rm{(k)}}) \,.
\end{align*}

\textbf{ii) Adjustment step.} We recompute the variational distribution $\rhoVB$ solution to \eqref{eq:final_min_max} for the decision $\hat\delta^{\rm{(k+1)}}$. The pseudo-code of our general method (denoted as \algVB) is shown in \cref{alg:alg_VB}. In the following section, we will introduce specific statistical models to which this algorithm can be applied.
\begin{algorithm}
\caption{\algVB: Portfolio Construction with Variational Bayes.}
\label{alg:alg_VB}
\textbf{Input:} Dataset $H_n$, Prior $\pi_0$ on $\theta$, initial decision guess $\hat\delta^{(0)}$, decision space $\cD$.\\
\For{$k=1, \dots, $}{
\While{Not converging}{
$\rhoVB\gets T(\rhoVB)$ where $T$ is the fixed-point operator defined in \cref{lemma:rho_GW} for GW model, \cref{lemma:rho_Ar} for AR model, \cref{lemma:rho_GP} for GP model.}
$\hat\delta^{(k+1)}\gets \mathrm{Proj}_{\cD}\left(\hat\delta^{(k)} - \alpha_k \nabla_\delta \obj(\hat\delta^{(k)})\right)$ , where $\obj$ is defined in \cref{lemma:objective_function_standard_GW} for GW model, \cref{lemma:objective_function_AR1} for AR model and \cref{lemma:objective_function_GP} for GP model.}
Return $\hat\delta^{(\infty)} = \deltaVB$.
\end{algorithm}

\textbf{Convexity properties of the objective function.} An important property of the objective function \eqref{eq:objective_function} is that it enjoys remarkable properties such as \emph{convexity} and \emph{smoothness}. Therefore, applying Projected Gradient Descent on $\delta\mapsto\obj(\delta)$, where the projection set $\cD$ is compact and convex ensures that the iterates $(\hat\delta^{(k)})_k$ will converge to an optimal point with value $\obj(\deltaVB)$, that is, $\obj(\hat\delta^{(k)})\to_k \obj(\deltaVB)$ at rate $\mathcal{O}(1/k)$. We state formally these results in \cref{prop:convexity_of_psi}. These properties will also play a crucial role in establishing the statistical convergence of $\deltaVB$.

\textbf{Statistical guarantees of Variational Bayes.} The restriction of the variational formulation to a smaller set of measures introduces a bias in the resulting decisions. There is a growing literature studying the statistical properties of variational Bayes approximation \citep{alquier2016properties,wang2019frequentist,alquier2020concentration,yang2020alpha,ray2022variational,huix2024theoretical}. Those results are not directly transferable to our problem because we do not only require the convergence of the approximate measure but the convergence with respect to the $\argmin$ of the objective function; we show that our VB algorithm converges asymptotically with respect to the sample size $n$ (see \cref{sec:theoretical_guarantees}).
\section{Application to Specific Statistical Models}
\label{sec:algorithm}
So far, \cref{alg:alg_VB} remained theoretical since we do not introduce assumptions on the statistical model $P_{\theta^*}$ yet, \emph{i.e.} we derived a general algorithm that holds for any parametric statistical model $P_{\theta^*}$. We now introduce relevant statistical models in the context of finance, where the core problem is to estimate the mean of investment returns, and the correlation between these returns. For all these models, we derive the corresponding fixed-point operator and the objective function $\mathcal{R}_{\Fmf}$ in closed form in \cref{sec:app_complete_statements}.
\subsection{Gaussian-Wishart (GW)}
\label{subsec:standard_Gaussian_Wishart}
For this model, observations (returns) are assumed independent and Normally distributed with unknown mean $\mu$ and precision $\Lambda$; putting a Gaussian prior on the mean and a Wishart prior on the precision matrix,
\begin{align}\label{eq:model_GW}
&Y_t \condi \mu, \Lambda \overset{\rm{i.i.d.}}{\sim} \cN(\mu, \Lambda^{-1})&&\forall t\in \mathbb{N}\nonumber \\
& \mu\sim\cN(\mu_0, \Lambda_0^{-1})\,,\quad\Lambda\sim\cW(\nu_0, \psi_0)\,.
\end{align}
Since we put prior on both mean and covariance $\theta  =(\mu, \Lambda)$, this model does not have closed-form moments for its \emph{joint} posterior distribution $\pi(\dint(\mu, \Lambda)\condi H_n)$, so we cannot compute the integral \eqref{eq:optimal_Bayesian_decision_exp} directly. The full expression of the fixed-point operator in \cref{prop:fixed_point} for this model is derived in \cref{lemma:rho_GW}, along with the corresponding objective function $\obj$ detailed in \cref{lemma:objective_function_standard_GW}.

\subsection{Autoregressive Model (AR)}
The non-dynamic model defined in \cref{eq:model_GW} is rather conservative, as it assumes no autocorrelation in returns, treating them as independent across time. While this simplifies learning and estimation, it overlooks the temporal dependencies often present in financial data such as market trends. Ignoring these patterns may limit the model's capacity to capture the true structure of returns. To circumvent this limitation, we introduce a model that incorporates a dynamic in the observations. We first outline a few definitions.
\begin{definition}[Matrix normal distribution \citep{quintana1987multivariate}]
\label{def:matrix normal distribution}
We say that $X\in\real^{d\times d}$ follows a matrix normal distribution with mean parameter $M\in\real^{d\times d}$, row-variance $U\in\real^{d\times d}$ and column variance $V\in\real^{d\times d}$ and denote $X\sim \cMN(M, U, V)$ if and only if
\begin{align*}
    \ve(X)\sim\cN(\ve(M), V\otimes U)\,,
\end{align*}
where we define $\ve(A)$ as the concatenated vector in $\real^{mn}$ of a matrix $A\in\real^{m\times n}$, $\ve(A) = (A_1,\cdots A_n)$. 
\end{definition}
In this model, we arbitrarily\footnote{This first observation can be set thanks to previously collected data, which may be available in practice.} set the initial value $Y_0 \sim \delta_y$ with $y\in \real^d$. Then, the model writes
\begin{align}\label{eq:model_AR1}
    &Y_t \condi Y_{t-1}, \Gamma, \Lambda \sim \cN(\Gamma Y_{t-1}, \Lambda)\qquad \forall t \in \mathbb{N}^*\nonumber\\
    &\Gamma \sim \cMN(M_0, U_0, V_0)\,,\quad\Lambda\sim\cW(\nu_0, \psi_0)\,.
\end{align}
The full expression of the fixed-point operator in \cref{prop:fixed_point} for this model is derived in \cref{lemma:rho_Ar}, along with the corresponding objective function $\obj$ detailed in \cref{lemma:objective_function_AR1}.

\subsection{Gaussian Process Model (GP)}
Gaussian processes \citep[GPs;][]{williams2006gaussian} can model the correlations between returns without assuming a specific functional form, making them particularly well-suited for environments with non-linear dependencies. We first define formally multivariate GPs.

\begin{definition}[Multivariate Gaussian process (MGP) \citep{chen2020multivariate}]
$f$ follows a multivariate Gaussian process with mean function $\mu:\real\to\real$, row variance function $k:\real^2\to\real$ and column variance $\Omega$, and we denote $f(\cdot)\sim\MGP(\mu(\cdot), k(\cdot, \cdot),\Omega)$, if, for every set of points $\{1, \dots, m\}\,$ with m any integer, we have $\left(f(t_1)^\top,\dots,f(t_m)^\top\right)^{\top} \sim \cMN(M_0^m, \Sigma_0^m, \Omega)$, where $[M_0^m]_{ij}= \mu(t_i)_j$ and $[\Sigma_0^m]_{ij} = k(t_i, t_j)$.
\end{definition}
Putting a multivariate GP prior on the mean returns $\mu$, the GP model is defined as
\begin{align}
    \label{eq:model_GP}
    &Y_t\condi \mu(\cdot), \Lambda \sim \cN(\mu(t), \Lambda^{-1})\qquad\forall t \in \mathbb{N}\\
    &\mu(\cdot) \sim\MGP(\mu_0(\cdot), K_0(\cdot), \Omega_0),\,\quad\Lambda\sim\cW(\nu_0, \psi_0)\,,\nonumber
\end{align}
where we emphasise that at time step $t$, $\mu(t)\in \real^d$ (\emph{i.e.} $(Y_t)_{t\geq 1}$ is a multivariate stochastic process). We will use specific kernel functions $K_0$ in numerical experiments. The full expression of the fixed-point operator in \cref{prop:fixed_point} for this model is derived in \cref{lemma:rho_GP}, and the corresponding objective function $\obj$ in \cref{lemma:objective_function_GP}.

\begin{remark}[Computing the gradient $\nabla_\delta\obj$]
For the objective functions presented in \cref{lemma:objective_function_standard_GW,lemma:objective_function_AR1,lemma:objective_function_GP}, we use automatic differentiation techniques to compute their gradients \citep{jax2018github}.
\end{remark}
\section{Numerical Experiments}
\label{sec:experiments}
\subsection{Experiments on Real-world Dataset}
\textbf{Dataset.} We use financial indices associated with the G20 member countries, spanning the period from 2012 to 2024; these data are publicly available\footnote{\url{https://finance.yahoo.com/markets/world-indices/}}. These indices are chosen over individual stock prices to minimize selection and survivorship biases. We apply Exponential Moving Averages (EMA) \citep{brockwell2002introduction} with 8 different scales to each index and compute the corresponding EMA for all indices. The EMA-transformed signals are then aggregated by averaging across scales, producing dataset with $d=8$ experts, where each column represents the averaged EMA signal at a specific scale, capturing smoothed trends across the indices. Monthly observations are extracted from this transformed dataset, yielding three different settings of increasing sample sizes: $(n, d) = (12, 8)$ \textbf{(Setting 1)}, $(n, d) = (48, 8)$ \textbf{(Setting 2)}, and $(n, d) = (84, 8)$ \textbf{(Setting 3)}.

\textbf{Baselines.} We compare our algorithm, \algVB, instantiated with models \eqref{eq:model_GW}, \eqref{eq:model_AR1} and \eqref{eq:model_GP} against several baseline portfolios.  The first is the Equal Weights portfolio (\textbf{EW}, also called the 1/$d$ portfolio), which assigns uniform weights to all assets: $\hat{\delta}_{\rm{EW}} = \frac{1}{d} 1_d$. We also consider the Markowitz portfolio (\textbf{Mwz}), as described in \cref{eq:Markovitz}, $\hat{\delta}_{\rm{Mwz}} = \frac{1}{\lambda} \hat{\Sigma}_n^{-1} \hat{\mu}_n$. Additionally, a more refined approach is to regularize the covariance matrix estimate, which is particularly advantageous in data-poor regimes. The Ledoit-Wolf method \citep{ledoit2003improved} employs a \emph{shrinkage} technique to stabilize the sample covariance matrix by combining it with a structured target matrix, typically a scaled identity matrix. The resulting shrunk covariance matrix is defined as $\hat{\Sigma}_n^{\rm{LW}} = (1 - \alpha)\hat{\Sigma}_n + \alpha I_d$, where $\alpha$ is the shrinkage intensity. Notably, the optimal value of $\alpha$ can be explicitly computed, as derived in \citet{ledoit2003improved}. This adjustment balances bias and variance, resulting in a better-conditioned estimator for high-dimensional settings. We define the corresponding Markowitz-based portfolio as $\hat{\delta}_{\rm{Mtz}}^{\rm{LW}} = \frac{1}{\lambda} \left(\hat{\Sigma}_n^{\rm{LW}}\right)^{-1} \hat{\mu}_n$. Finally, we compare our approach against a state-of-the-art method, the Exponential Utility for Gaussian Mixtures (\textbf{EGM}), which maximizes the exponential utility under a Gaussian mixture model assumption for returns \citep{luxenberg2024portfolio}.

\textbf{Prior parameters} For all models, the Wishart prior is set as $\nu_0 = d$ and $\psi_0  = \frac{1}{\nu_0}\hat\Sigma_n^{-1}$, where $\hat\Sigma_n$ is the empirical estimate of covariance matrix. For the GW model, we set the prior mean as the empirical mean, $\mu_0 = \hat\mu_n$ and $\Sigma_0 = I_d$.  For the AR model, we set $M_0$ as the MLE estimate of the transition matrix $\Gamma$ obtained via linear regression on the observations $H_n$, $M_0 = \left(\sum_{t=1}^n Y_t Y_{t-1}^\top\right)\left( \sum_{t=1}^n Y_t Y_t^\top \right)^{-1}$. We set the row-covariance $U_0$ and column-covariance $V_0$ as identities. For the GP model \eqref{eq:model_GP}, we choose a Radial basis function kernel parameterized by $\gamma$ \emph{i.e.} 
$\forall t_1, t_2,\,\quad k_\gamma(t_1, t_2) = \exp\left(\frac{(t_1 - t_2)^2}{2\gamma^2}\right)$,
and tune $\gamma$ trough Gradient-based optimization (with respect to marginal likelihood). We set the mean function to $0$ and the prior column variance as $\Omega_0 = I_d$. The hyperparameter choices are discussed in \cref{sec:app_additional_experiments}.

\begin{figure*}[t]  
    \centering
    \includegraphics[width=17cm]{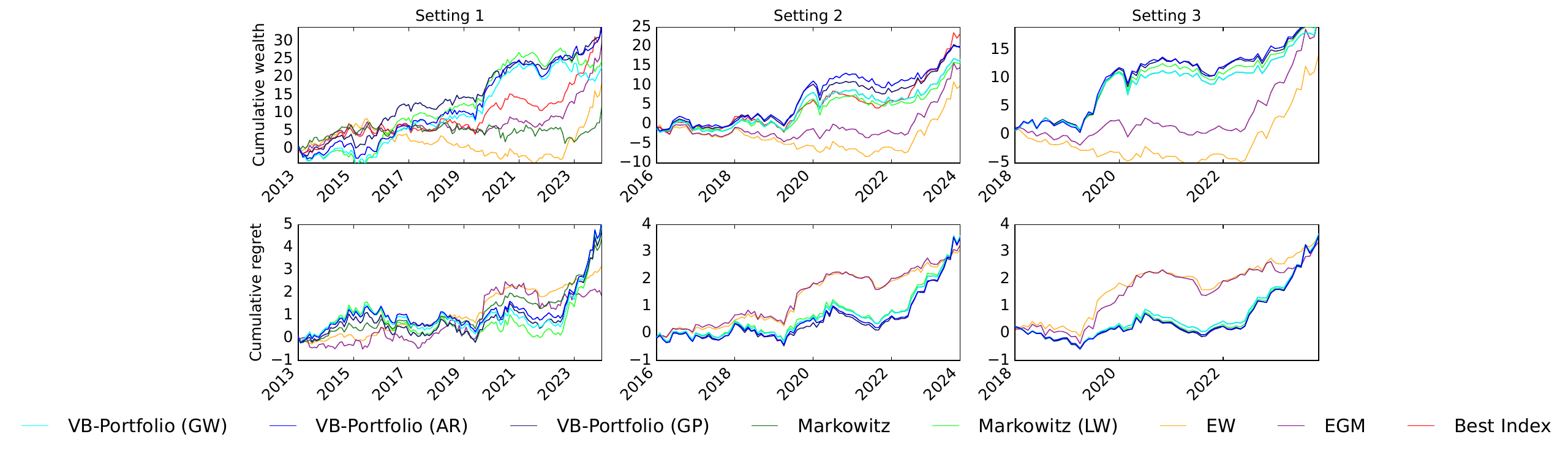}
    \caption{Cumulative wealth (row 1) and cumulative regret with respect to best index in hindsight (row 2) in 3 settings. For each strategy, we rescale the cumulative plot by the standard deviation of the returns.}
    \label{fig:wealth_and_regret}
\end{figure*}

\textbf{Results (Cumulative wealth and regret).}
For each allocation strategy, we compute the out-of-sample cumulative wealth, $\delta^\top\sum_{y\in\mathcal{E}_{\rm{test}}}y$ ($\mathcal{E}_{\rm{test}}$ are observations of the testing set, \emph{i.e.} observations from 2013 for Setting 1, from 2016 for Setting 2 and from 2018 for Setting 3). To enable a fair comparison across strategies with varying levels of risk, we rescale each cumulative wealth by its standard deviation. This risk-adjusted rescaling is a standard convention in portfolio construction literature. Additionally, we plot the strategy corresponding to allocating all mass on the \emph{best index in hindsight}, defined as the index with the highest cumulative wealth at the end of the testing horizon. To further assess performance, we plot the \emph{cumulative regret} for each strategy against this best index in hindsight, that is, the difference between the cumulative wealth of the best index in hindsight and the one of the given strategy. We rescale this cumulative difference by its standard deviation. Results are displayed in \cref{fig:wealth_and_regret}, and show that while \algVB(GW) exhibits performance comparable to the Markowitz-based portfolio, $\algVB$, when instantiated with both the AR and GP models, outperforms the other strategies overall. This superior performance suggests that these models are particularly effective at adapting to evolving market conditions.

\textbf{Sharpe Ratios Comparison}
\begin{table} 
    \centering
    \caption{Out-of-sample annualized Sharpe ratios of each portfolio for different settings.}
    \label{tab:sharpe_ratios}
    \resizebox{0.5\textwidth}{!}{ 
    \begin{tabular}{@{}lccc@{}}
        \toprule
        \multirow{2}{*}{\textbf{Allocation Strategy}} & \multicolumn{3}{c}{\textbf{Annualized Sharpe Ratio}} \\
        \cmidrule(lr){2-4}
         & Setting 1 & Setting 2 & Setting 3 \\
        \midrule
        \texttt{algVB (GW)} & $0.59$ & $0.77$ & $1.03$ \\
        \texttt{algVB (AR)} & $0.88$ & \textbf{0.90} & \textbf{1.16} \\
        \texttt{algVB (GP)} & \textbf{0.90} & \textbf{0.90} & $1.12$ \\
        \texttt{Markowitz} & $0.31$ & $0.77$ & $1.03$ \\
        \texttt{Markowitz (LW)} & $0.63$ & $0.77$ & $1.11$ \\
        \texttt{Equal Weights} & $0.52$ & $0.52$ & $0.74$ \\
        \texttt{EGM} & $0.80$ & $0.80$ & $1.05$ \\
        \bottomrule
    \end{tabular}%
    }
\end{table}

The Sharpe ratio \citep{sharpe1966mutual,sharpe1994sharpe} is a widely used metric for assessing the risk-adjusted performance of investment strategies. It is defined as the ratio of the mean return to the standard deviation, quantifying the return per unit of risk. Let $\hat\mu_{\rm{test}}(\delta)$ denotes the mean return of strategy $\delta$ over the testing set, and $\hat\sigma_{\rm{test}}(\delta)$ its standard deviation. The annualized Sharpe ratio is then computed as $\rm{SR}(\delta) = \sqrt{12}\hat\mu_{\rm{test}}(\delta) / \hat\sigma_{\rm{test}}(\delta)$. This metric facilitates meaningful comparisons across strategies by highlighting those that deliver higher returns relative to risk. As illustrated in \cref{tab:sharpe_ratios}, \algVB(AR) achieves the highest Sharpe ratio overall, closely followed by \algVB(GP). Both methods consistently outperform traditional approaches, such as the Markowitz portfolio, particularly in Settings 1 and 2, where the smaller sample sizes lead to less stable estimates for the other strategies. These findings underscore the robustness of the proposed methods in data-poor environments, demonstrating that Bayesian approaches are well-suited to regularizing estimates and mitigating the impact of limited training data.

\subsection{Numerical Consistency} 
To assess the consistency of our approximation $\deltaVB$ to $\delta^*$, we use a gradient descent-based algorithm that leverages Markov Chain Monte Carlo (MCMC) sampling to estimate the gradient of the objective function.

\textbf{Approximating the objective function.} First, we want to approximate the integral \eqref{eq:optimal_Bayesian_decision_exp} for any $\delta\in\cD$;
\begin{align*}
     \mathcal{R}_{\cMplus}(\delta) = \int_{\mathcal{Y}} e^{-\lambda\delta^\top \Ypred}\pi(\dint\Ypred \condi H_n) \approx \frac{1}{M}\sum_{m=1}^M e^{-\lambda \delta^\top y^{(k)}}\,,
\end{align*}
where $(y^{(k)})_{k\in[M]}$ are $M$ samples from the predictive posterior distribution $\pi(\cdot\condi H_n)$. This can be done with the Gibbs sampling algorithm \citep{geman1984stochastic}: in fact, by remarking that
\begin{align*}
    \pi(\dint\Ypred\condi H_n) = \int_{\Theta}\pi(\dint\Ypred\condi\theta)\pi(\dint\theta\condi H_n)\,,
\end{align*}
and since we now how to sample from the \emph{conditional posterior} $\pi(\dint\theta_i\condi \bm{\theta}_{-i}, H_n)$, we can generate $M$ samples $(\theta^{(m)})_{m\leq M}$ from the joint posterior distribution $\pi(\dint\theta\condi H_n)$. From this sequence, we can now sample from the distribution $\Breve{\pi}(\dint \Ypred)\propto e^{-\lambda\delta^\top \Ypred} \pi(\dint\Ypred \mid H_n)$ \emph{conditionally} on one sample $\theta^{(k)}$, by drawing a sample $y^{(k)}$ from the distribution
\begin{align*}
\Breve{\pi}_k(\dint\Ypred)\propto e^{-\lambda^\top \Ypred}\pi(\dint\Ypred;\theta^{(k)})\,,
\end{align*}
resulting in a sequence of $M$ samples $(y^{(k)})_{k\in[M]}$. 

\textbf{Approximating the gradient $\nabla_\delta\mathcal{R}$.} By Leibniz rule, we have
\begin{align*}
\nabla_{\delta}\mathcal{R}_{\cMplus}(\delta) &= - \lambda \int_{\mathcal{Y}}\Ypred \frac{e^{-\lambda\delta^\top \Ypred}\pi(\dint\Ypred \mid H_n)}{\int_{\mathcal{Y}} e^{-\lambda\delta^\top \Ypred} \pi(\dint\Ypred \mid H_n)} = -\lambda \E{\Breve{\pi}}{\Ypred}\,,
\end{align*}
for which we can approximate by
\begin{align*}
    \nabla_{\delta}\mathcal{R}_{\cMplus}(\delta) \approx -\lambda \frac{1}{M}\sum_{k=1}^M z^{(k)}\,,\quad\text{ where }z^{(k)}\sim \Breve{\pi}_k(\cdot)\,.
\end{align*}

The pseudo-code of this MCMC algorithm is shown in \cref{alg:MCMC_GW}. We instantiate this algorithm for GW and AR model in \cref{app:MCMC_specific}, where we provide in particular the expressions of the conditional posteriors and $\check{\pi}$.
\begin{algorithm}
\caption{\MCMC: Portfolio Construction with Markov Chain Monte-Carlo.}
\label{alg:MCMC_GW}
\textbf{Input:} Dataset $H_n$, initial decision $\hat\delta^{(0)}$, number of Monte-Carlo samples $M$, risk parameter $\lambda$, step-size $\eta$. \\
\While{Not converging}{Get $M$ samples $(\theta^{(k)})_{k\in[M]}$ from Gibbs sampler.\\
For all $k \in [M]$, sample $z^{(k)} \sim \Breve{\pi}_k$.\\
$\hat\delta^{(k+1)} \gets \mathrm{Proj}_{\cD}\left(\hat\delta^{(k)} + \eta\lambda\frac{1}{M} \sum_{k \in [M]} z^{(k)}\right)$}
Return $\hat\delta^{(\infty)} = \hat\delta^{\rm{MCMC}}$.
\end{algorithm}

\textbf{Consistency with synthetic data.}  
\begin{figure}[bth]
    \centering
    \includegraphics[width=0.3\linewidth]{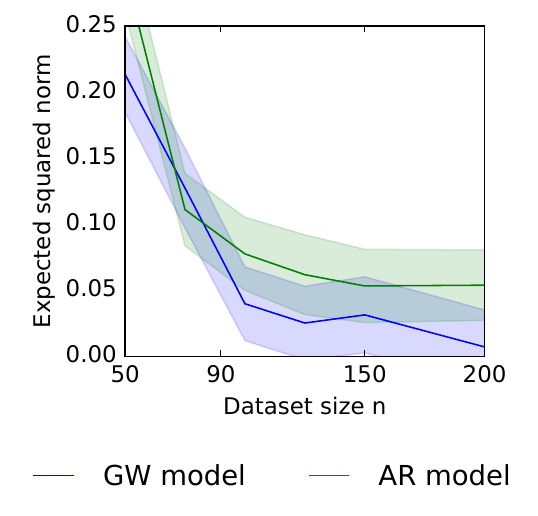}
    \caption{Expected 2-norm $\E{}{\lVert \hat\delta^{\rm{MCMC}} - \deltaVB \rVert_2}$ as a function of dataset size $n$ for both the GW and AR(1) models. Each point represents the average over 50 iterations, with error bars indicating one standard deviation to represent the confidence interval. The dimensionality of the data is fixed at $d=30$.}
    \label{fig:consistency}
\end{figure}
To evaluate the numerical consistency of our method, we show that $\E{}{\lVert \hat\delta^{\rm{MCMC}} - \deltaVB \rVert_2}$ converges to $0$ as the sample size $n$ grows. 
We generate synthetic datasets for both GW model and AR model with $d=30$. For the GW model, we randomly set each component of the true mean return, $\mu^*_i$, according to a uniform distribution, $\mu^*_i \sim \mathcal{U}([0, 1])$, and use an identity covariance matrix, $\Sigma^* = I_d$. For the AR model, the true transition matrix $\Gamma^*$ is set to a diagonal matrix with values evenly spaced from 0.6 to 0.99, while the covariance matrix is $\Sigma^* = 0.1 I_d$. We generate datasets of varying sizes, with $n$ ranging from 50 to 200. For the GW model, we simulate data according to \eqref{eq:model_GW}, and for the AR model, we follow \eqref{eq:model_AR1}. For each dataset, we compute the decision vector using both \algVB~ and \MCMC, where the Gibbs sampler is run for $M=20{,}000$ iterations. We compute $\E{}{\lVert \hat\delta^{\rm{MCMC}} - \deltaVB \rVert_2}$, where the expectation is taken over 50 repetitions with newly generated datasets.

\cref{fig:consistency} illustrates the relationship between this expected norm difference and the dataset size $n$. As $n$ increases, we observe that the difference between the two decision methods diminishes, confirming the numerical consistency of \algVB~with respect to \MCMC as the sample size grows.
\section{Theoretical Guarantees}
\label{sec:theoretical_guarantees}
In this section, we assume that $\cD$ is compact and convex, as is the case when $\cD = \Delta_d$, the standard simplex. As discussed earlier, we cannot directly rely on existing results concerning the statistical efficiency of variational Bayes (VB) approximations. Instead, we focus on guarantees on $\deltaVB$ which is a critical point of the objective function of the VB approximation. We introduce the notation $\hat{\delta}^{(k)}$ for the result of our algorithm after $k$ iterations of \cref{alg:alg_VB}. Under the formal conditions stated below, we establish the following two key theoretical results: 

\textbf{Numerical convergence.} $\obj(\hat{\delta}^{(k)})$ converges to $\obj(\deltaVB)$ with respect to the number of iterations $k$ at rate $\mathcal{O}(1/k)$.

\textbf{Statistical consistency.} $\deltaVB$ converges with respect to the sample size $n$ to a Markowitz decision.

\subsection{Numerical Convergence} We begin by addressing the convergence of our algorithm, which relies on the convexity and smoothness properties of the objective function $\mathcal{R}_{\Fmf}$.
\begin{proposition}[Properties of $\mathcal{R}_{\Fmf}$]\label{prop:convexity_of_psi}
For a fixed dataset size $n$, $\delta\mapsto\obj(\delta)$ is convex and smooth. Let $L$ denote the smoothness constant of $\mathcal{R}_{\Fmf}$. Using Gradient descent with a step size of $\eta = 1/L$ and initial point $\hat\delta^{(0)}$ ensures
\begin{align*}
    \obj(\hat\delta^{(k)}) - \obj(\deltaVB)\leq \frac{2L}{k-1}\lVert \hat\delta^{(0)} - \deltaVB\rVert_2\,.
\end{align*}
\end{proposition}
This result shows that if the fixed-point iteration converges to $\rhoVB$ at each step we compute $\hat\delta^{(k)}$, then $\obj(\delta^{(k)})$ converges to $\obj(\deltaVB)$ at a rate $\mathcal{O}(1/k)$. The proof of \cref{prop:convexity_of_psi} (given in \cref{subsec:app_proof_convecity}) relies on expressing $\obj$ as a Fenchel-Legendre transformation of the \emph{strongly} convex map $\rho\mapsto\KL(\rho, \pi_n)$.

\subsection{Statistical Consistency}
Next, we establish asymptotic consistency results as the sample size $n\to+\infty$, focusing on the behaviour of the \emph{decision} $\deltaVB$. For this, we introduce an assumption regarding the asymptotic behaviour of the fixed-point equation.
\begin{assumption}\label{assumption:contractance}
As the sample size $n\to+\infty$, the variational distribution converges pointwise,
\begin{align*}
    \forall \delta\in\cD\,,\quad \rhoVB(\delta)\xrightarrow{n\to+\infty} \hat\rho_\infty(\delta),
\end{align*}
where $\hat\rho_\infty$ is solution to the asymptotic fixed-point operator: denoting $T_n$ the fixed point operator defined in \cref{prop:fixed_point}, for any $\delta\in\cD$, if $T_n(\rhoVB(\delta)) = \rhoVB(\delta)$ then the operator $T_\infty = \lim_{n\to+\infty} T_n$ satisfies $T_\infty( \hat\rho_\infty(\delta)) =  \hat\rho_\infty(\delta)$.
\end{assumption}
\cref{assumption:contractance} is a relatively strong assumption, as rigorously proving it would require showing that the fixed-point operator is \emph{contractant} across all models considered. However, we provide numerical evidence in \cref{ref:assumption_comments} to support this assumption, indicating that it is reasonable in practice.
Next we derive asymptotic consistency of the variational decision $\deltaVB$ under this assumption.

\begin{theorem}[Consistency of the variational decision]\label{th:consistency}
Under both \emph{GW} \eqref{eq:model_GW} and \emph{AR} \eqref{eq:model_AR1} models, the variational decision converges almost surely to the Markovitz decision in $\cD$,
\begin{align*}
    \deltaVB \xrightarrow[n\to+\infty]{a.s.}\argmin_{\delta\in\cD}\left\{\frac{1}{2}\delta^\top \hat\Sigma_\infty^{-1}\delta - \lambda\delta^\top \hat\mu_\infty \right\} \,.
\end{align*}
where $\hat\Sigma_\infty$ and $\hat\mu_\infty$ are the  empirical estimates of the covariance matrix and mean vector, respectively, in the limit as $n\to+\infty$.
\end{theorem}
The proof of \cref{th:consistency}, given in \cref{subsec:app_proof_consistency}, involves a key technical challenge: interchanging the limit and the $\argmin$
. Specifically, we need to show:
\begin{align*}
    \lim_{n\to+\infty}\argmin_{\delta\in\cD}\obj(\delta) = \argmin_{\delta\in\cD}\lim_{n\to+\infty}\obj(\delta)\,.
\end{align*}
In particular, this step requires \emph{epi-convergence} of the sequence $(\obj)_n$, which can be leveraged by the compacity property of $\Fmf$ and $\cD$. 
\section{Conclusion}
We showed that Bayesian optimal decision for exponential utility can be interpreted as a saddle-point problem. We developed a computationally efficient algorithm based on variational Bayes with provable convergence guarantees, demonstrating its effectiveness in real-world portfolio optimization problems.

\textbf{Maximizing exponential utility functions.} Our min-max formulation (\cref{th:main_th}) provides a versatile framework for scenarios where the expected utility lacks a closed-form solution. This methodology not only bridges theoretical and practical domains but also holds promise for broader applications, particularly in areas like reinforcement learning \citep{marthe2024beyond}, where exponential utility functions are pivotal for navigating decision-making under uncertainty.

\textbf{Beyond Gradient-Descent.}Although our objective function is convex and smooth, leveraging advanced optimization techniques could unlock further potential. Techniques such as Nesterov’s acceleration \citep{nesterov2018lectures} and mirror-descent-based methods \citep{nemirovski2004prox} for saddle-point optimization present opportunities to enhance convergence rates and scalability. These methods could prove especially beneficial for portfolio construction in high-dimensional settings.
\section{Acknowledgements}
The authors would like to warmly thank Emmanuel Sérié for insightful discussions and remarks on this work. We also thank Vincent Fortuin for feedbacks.

Nicolas Nguyen and Claire Vernade are funded by the Deutsche Forschungsgemeinschaft (DFG) under both the project 468806714 of the Emmy Noether Programme and under Germany’s Excellence Strategy – EXC number 2064/1 – Project number 390727645. 

Nicolas Nguyen and Claire Vernade thank the international Max Planck Research School for Intelligent Systems (IMPRS-IS).
\bibliography{references} 

\begin{thebibliography}{51}
\providecommand{\natexlab}[1]{#1}
\providecommand{\url}[1]{\texttt{#1}}
\expandafter\ifx\csname urlstyle\endcsname\relax
  \providecommand{\doi}[1]{doi: #1}\else
  \providecommand{\doi}{doi: \begingroup \urlstyle{rm}\Url}\fi

\bibitem[Agrawal et~al.(2022)Agrawal, Roy, and Uhler]{agrawal2022covariance}
Raj Agrawal, Uma Roy, and Caroline Uhler.
\newblock Covariance matrix estimation under total positivity for portfolio selection.
\newblock \emph{Journal of Financial Econometrics}, 20\penalty0 (2):\penalty0 367--389, 2022.

\bibitem[Alquier(2024)]{alquier2024user}
Pierre Alquier.
\newblock User-friendly introduction to pac-bayes bounds.
\newblock \emph{Foundations and Trends{\textregistered} in Machine Learning}, 17\penalty0 (2):\penalty0 174--303, 2024.

\bibitem[Alquier and Ridgway(2020)]{alquier2020concentration}
Pierre Alquier and James Ridgway.
\newblock Concentration of tempered posteriors and of their variational approximations.
\newblock \emph{The Annals of Statistics}, 2020.

\bibitem[Alquier et~al.(2016)Alquier, Ridgway, and Chopin]{alquier2016properties}
Pierre Alquier, James Ridgway, and Nicolas Chopin.
\newblock On the properties of variational approximations of gibbs posteriors.
\newblock \emph{The Journal of Machine Learning Research}, 17\penalty0 (1):\penalty0 8374--8414, 2016.

\bibitem[Bach(2024)]{bach2024learning}
Francis Bach.
\newblock \emph{Learning theory from first principles}.
\newblock MIT press, 2024.

\bibitem[Barry(1974)]{barry1974portfolio}
Christopher~B Barry.
\newblock Portfolio analysis under uncertain means, variances, and covariances.
\newblock \emph{The Journal of Finance}, 29\penalty0 (2):\penalty0 515--522, 1974.

\bibitem[Benaych-Georges et~al.(2023)Benaych-Georges, Bouchaud, and Potters]{benaych2023optimal}
Florent Benaych-Georges, Jean-Philippe Bouchaud, and Marc Potters.
\newblock Optimal cleaning for singular values of cross-covariance matrices.
\newblock \emph{The Annals of Applied Probability}, 33\penalty0 (2):\penalty0 1295--1326, 2023.

\bibitem[Benichou et~al.(2016)Benichou, Lemp{\'e}ri{\`e}re, Kockelkoren, Seager, Bouchaud, Potters, et~al.]{benichou2016agnostic}
Raphael Benichou, Yves Lemp{\'e}ri{\`e}re, Julien Kockelkoren, Philip Seager, Jean-Philippe Bouchaud, Marc Potters, et~al.
\newblock Agnostic risk parity: Taming known and unknown-unknowns.
\newblock \emph{arXiv preprint arXiv:1610.08818}, 2016.

\bibitem[Billingsley(2013)]{billingsley2013convergence}
Patrick Billingsley.
\newblock \emph{Convergence of probability measures}.
\newblock John Wiley \& Sons, 2013.

\bibitem[Bishop(2006)]{bishop2006pattern}
C~Bishop.
\newblock Pattern recognition and machine learning.
\newblock \emph{Springer google schola}, 2:\penalty0 531--537, 2006.

\bibitem[Black and Litterman(1992)]{black1992global}
Fischer Black and Robert Litterman.
\newblock Global portfolio optimization.
\newblock \emph{Financial analysts journal}, 48\penalty0 (5):\penalty0 28--43, 1992.

\bibitem[Boyd and Vandenberghe(2004)]{boyd2004convex}
Stephen Boyd and Lieven Vandenberghe.
\newblock \emph{Convex optimization}.
\newblock Cambridge university press, 2004.

\bibitem[Bradbury et~al.(2018)Bradbury, Frostig, Hawkins, Johnson, Leary, Maclaurin, Necula, Paszke, Vander{P}las, Wanderman-{M}ilne, and Zhang]{jax2018github}
James Bradbury, Roy Frostig, Peter Hawkins, Matthew~James Johnson, Chris Leary, Dougal Maclaurin, George Necula, Adam Paszke, Jake Vander{P}las, Skye Wanderman-{M}ilne, and Qiao Zhang.
\newblock {JAX}: composable transformations of {P}ython+{N}um{P}y programs, 2018.
\newblock URL \url{http://github.com/google/jax}.

\bibitem[Brockwell and Davis(2002)]{brockwell2002introduction}
Peter~J Brockwell and Richard~A Davis.
\newblock \emph{Introduction to time series and forecasting}.
\newblock Springer, 2002.

\bibitem[Bun et~al.(2017)Bun, Bouchaud, and Potters]{bun2017cleaning}
Jo{\"e}l Bun, Jean-Philippe Bouchaud, and Marc Potters.
\newblock Cleaning large correlation matrices: tools from random matrix theory.
\newblock \emph{Physics Reports}, 666:\penalty0 1--109, 2017.

\bibitem[Cesa-Bianchi and Lugosi(2006)]{cesa2006prediction}
Nicolo Cesa-Bianchi and G{\'a}bor Lugosi.
\newblock \emph{Prediction, learning, and games}.
\newblock Cambridge university press, 2006.

\bibitem[Chen et~al.(2020)Chen, Wang, and Gorban]{chen2020multivariate}
Zexun Chen, Bo~Wang, and Alexander~N Gorban.
\newblock Multivariate gaussian and student-t process regression for multi-output prediction.
\newblock \emph{Neural Computing and Applications}, 32:\penalty0 3005--3028, 2020.

\bibitem[Cover(1991)]{cover1991universal}
Thomas~M Cover.
\newblock Universal portfolios.
\newblock \emph{Mathematical finance}, 1\penalty0 (1):\penalty0 1--29, 1991.

\bibitem[De~Franco et~al.(2019)De~Franco, Nicolle, and Pham]{de2019bayesian}
Carmine De~Franco, Johann Nicolle, and Huy{\^e}n Pham.
\newblock Bayesian learning for the markowitz portfolio selection problem.
\newblock \emph{International Journal of Theoretical and Applied Finance}, 22\penalty0 (07):\penalty0 1950037, 2019.

\bibitem[Donsker and Varadhan(1983)]{donsker1983asymptotic}
Monroe~D Donsker and SR~Srinivasa Varadhan.
\newblock Asymptotic evaluation of certain markov process expectations for large time. iv.
\newblock \emph{Communications on pure and applied mathematics}, 36\penalty0 (2):\penalty0 183--212, 1983.

\bibitem[Geman and Geman(1984)]{geman1984stochastic}
Stuart Geman and Donald Geman.
\newblock Stochastic relaxation, gibbs distributions, and the bayesian restoration of images.
\newblock \emph{IEEE Transactions on pattern analysis and machine intelligence}, pages 721--741, 1984.

\bibitem[Gerber and Pafum(1998)]{gerber1998utility}
Hans~U Gerber and G{\'e}rard Pafum.
\newblock Utility functions: from risk theory to finance.
\newblock \emph{North American Actuarial Journal}, 2\penalty0 (3):\penalty0 74--91, 1998.

\bibitem[Harvey et~al.(2010)Harvey, Liechty, Liechty, and M{\"u}ller]{harvey2010portfolio}
Campbell~R Harvey, John~C Liechty, Merrill~W Liechty, and Peter M{\"u}ller.
\newblock Portfolio selection with higher moments.
\newblock \emph{Quantitative Finance}, 10\penalty0 (5):\penalty0 469--485, 2010.

\bibitem[Huix et~al.(2024)Huix, Korba, Durmus, and Moulines]{huix2024theoretical}
Tom Huix, Anna Korba, Alain Durmus, and Eric Moulines.
\newblock Theoretical guarantees for variational inference with fixed-variance mixture of gaussians.
\newblock \emph{arXiv preprint arXiv:2406.04012}, 2024.

\bibitem[Ismail and Pham(2019)]{ismail2019robust}
Amine Ismail and Huy{\^e}n Pham.
\newblock Robust markowitz mean-variance portfolio selection under ambiguous covariance matrix.
\newblock \emph{Mathematical Finance}, 29\penalty0 (1):\penalty0 174--207, 2019.

\bibitem[J{\'e}z{\'e}quel et~al.(2022)J{\'e}z{\'e}quel, Ostrovskii, and Gaillard]{jezequel2022efficient}
R{\'e}mi J{\'e}z{\'e}quel, Dmitrii~M Ostrovskii, and Pierre Gaillard.
\newblock Efficient and near-optimal online portfolio selection.
\newblock \emph{arXiv preprint arXiv:2209.13932}, 2022.

\bibitem[Jordan et~al.(1999)Jordan, Ghahramani, Jaakkola, and Saul]{jordan1999introduction}
Michael~I Jordan, Zoubin Ghahramani, Tommi~S Jaakkola, and Lawrence~K Saul.
\newblock An introduction to variational methods for graphical models.
\newblock \emph{Machine learning}, 37:\penalty0 183--233, 1999.

\bibitem[Kato(2024)]{kato2024general}
Masahiro Kato.
\newblock General bayesian predictive synthesis.
\newblock \emph{arXiv preprint arXiv:2406.09254}, 2024.

\bibitem[Ledoit and Wolf(2003)]{ledoit2003improved}
Olivier Ledoit and Michael Wolf.
\newblock Improved estimation of the covariance matrix of stock returns with an application to portfolio selection.
\newblock \emph{Journal of empirical finance}, 10\penalty0 (5):\penalty0 603--621, 2003.

\bibitem[Li and Hoi(2018)]{li2018online}
Bin Li and Steven Chu~Hong Hoi.
\newblock \emph{Online portfolio selection: principles and algorithms}.
\newblock Crc Press, 2018.

\bibitem[Luo et~al.(2018)Luo, Wei, and Zheng]{luo2018efficient}
Haipeng Luo, Chen-Yu Wei, and Kai Zheng.
\newblock Efficient online portfolio with logarithmic regret.
\newblock \emph{Advances in neural information processing systems}, 31, 2018.

\bibitem[Luxenberg and Boyd(2024)]{luxenberg2024portfolio}
Eric Luxenberg and Stephen Boyd.
\newblock Portfolio construction with gaussian mixture returns and exponential utility via convex optimization.
\newblock \emph{Optimization and Engineering}, 25\penalty0 (1):\penalty0 555--574, 2024.

\bibitem[Markowitz(1952)]{Markovitz1952}
Harry Markowitz.
\newblock Portfolio selection.
\newblock \emph{The Journal of Finance}, 7\penalty0 (1):\penalty0 77--91, 1952.

\bibitem[Marthe et~al.(2024)Marthe, Garivier, and Vernade]{marthe2024beyond}
Alexandre Marthe, Aur{\'e}lien Garivier, and Claire Vernade.
\newblock Beyond average return in markov decision processes.
\newblock \emph{Advances in Neural Information Processing Systems}, 36, 2024.

\bibitem[Merton(1969)]{merton1969lifetime}
Robert~C Merton.
\newblock Lifetime portfolio selection under uncertainty: The continuous-time case.
\newblock \emph{The review of Economics and Statistics}, pages 247--257, 1969.

\bibitem[Merton(1972)]{merton1972analytic}
Robert~C Merton.
\newblock An analytic derivation of the efficient portfolio frontier.
\newblock \emph{Journal of financial and quantitative analysis}, 7\penalty0 (4):\penalty0 1851--1872, 1972.

\bibitem[Nemirovski(2004)]{nemirovski2004prox}
Arkadi Nemirovski.
\newblock Prox-method with rate of convergence o (1/t) for variational inequalities with lipschitz continuous monotone operators and smooth convex-concave saddle point problems.
\newblock \emph{SIAM Journal on Optimization}, 15\penalty0 (1):\penalty0 229--251, 2004.

\bibitem[Nesterov et~al.(2018)]{nesterov2018lectures}
Yurii Nesterov et~al.
\newblock \emph{Lectures on convex optimization}, volume 137.
\newblock Springer, 2018.

\bibitem[Orabona and Jun(2023)]{orabona2023tight}
Francesco Orabona and Kwang-Sung Jun.
\newblock Tight concentrations and confidence sequences from the regret of universal portfolio.
\newblock \emph{IEEE Transactions on Information Theory}, 2023.

\bibitem[Parisi and Shankar(1988)]{parisi1988statistical}
Giorgio Parisi and Ramamurti Shankar.
\newblock Statistical field theory.
\newblock \emph{American Journal of Physics}, 1988.

\bibitem[Quinonero-Candela and Rasmussen(2005)]{quinonero2005unifying}
Joaquin Quinonero-Candela and Carl~Edward Rasmussen.
\newblock A unifying view of sparse approximate gaussian process regression.
\newblock \emph{The Journal of Machine Learning Research}, 6:\penalty0 1939--1959, 2005.

\bibitem[Quintana(1987)]{quintana1987multivariate}
Jose~Mario Quintana.
\newblock \emph{Multivariate Bayesian forecasting models}.
\newblock PhD thesis, University of Warwick, 1987.

\bibitem[Ray and Szab{\'o}(2022)]{ray2022variational}
Kolyan Ray and Botond Szab{\'o}.
\newblock Variational bayes for high-dimensional linear regression with sparse priors.
\newblock \emph{Journal of the American Statistical Association}, 117\penalty0 (539):\penalty0 1270--1281, 2022.

\bibitem[Robert(2007)]{robert2007bayesian}
Christian~P Robert.
\newblock \emph{The Bayesian choice: from decision-theoretic foundations to computational implementation}, volume~2.
\newblock Springer, 2007.

\bibitem[Rockafellar and Wets(2009)]{rockafellar2009variational}
R~Tyrrell Rockafellar and Roger J-B Wets.
\newblock \emph{Variational analysis}, volume 317.
\newblock Springer Science \& Business Media, 2009.

\bibitem[Sharpe(1966)]{sharpe1966mutual}
William~F Sharpe.
\newblock Mutual fund performance.
\newblock \emph{The Journal of business}, 39\penalty0 (1):\penalty0 119--138, 1966.

\bibitem[Sharpe(1994)]{sharpe1994sharpe}
William~F Sharpe.
\newblock The sharpe ratio.
\newblock \emph{Journal of portfolio management}, 21\penalty0 (1):\penalty0 49--58, 1994.

\bibitem[Van~Erven et~al.(2020)Van~Erven, Van~der Hoeven, Kot{\l}owski, and Koolen]{van2020open}
Tim Van~Erven, Dirk Van~der Hoeven, Wojciech Kot{\l}owski, and Wouter~M Koolen.
\newblock Open problem: Fast and optimal online portfolio selection.
\newblock In \emph{Conference on Learning Theory}, pages 3864--3869. PMLR, 2020.

\bibitem[Wang and Blei(2019)]{wang2019frequentist}
Yixin Wang and David~M Blei.
\newblock Frequentist consistency of variational bayes.
\newblock \emph{Journal of the American Statistical Association}, 114\penalty0 (527):\penalty0 1147--1161, 2019.

\bibitem[Williams and Rasmussen(2006)]{williams2006gaussian}
Christopher~KI Williams and Carl~Edward Rasmussen.
\newblock \emph{Gaussian processes for machine learning}, volume~2.
\newblock MIT press Cambridge, MA, 2006.

\bibitem[Yang et~al.(2020)Yang, Pati, and Bhattacharya]{yang2020alpha}
Yun Yang, Debdeep Pati, and Anirban Bhattacharya.
\newblock $\alpha$-variational inference with statistical guarantees.
\newblock \emph{The Annals of Statistics}, 48\penalty0 (2):\penalty0 886--905, 2020.

\end{thebibliography}
\newpage
\appendix
\section{Difference Between our Work and Online Portfolio Selection}
The field of machine learning has contributed to portfolio selection by framing it as an online learning problem \citep{cover1991universal,cesa2006prediction}. The OPS approach consists in sequentially allocating capital across assets to maximize the cumulative log-wealth over time. Formally, at each time step $t$, the investor selects a portfolio vector $\delta_t$ based on the information available up to that point, with the goal of maximizing the log-wealth achieved after $n$ rounds $\log\big(\prod_{t=1}^n R_t^\top \delta_t\big)$, where $(R_t)_t$ are the asset returns. This online framework has led to the development of a variety of algorithms with strong theoretical guarantees, particularly in terms of regret bounds, measuring how much worse the cumulative wealth of the algorithm is compared to that of an optimal strategy selected in hindsight. Early works such as Cover's universal portfolio algorithm \citep{cover1991universal} introduced a strategy that could asymptotically achieve the same wealth as the best constant-rebalanced portfolio. More recent advances have continued to refine these results, providing tighter regret bounds and more efficient learning mechanisms in both stochastic and adversarial settings \citep{li2018online,luo2018efficient,van2020open,jezequel2022efficient,orabona2023tight}.

Despite its theoretical richness and the mathematical sophistication, the OPS literature has seen limited adoption among practitioners. A key reason for this limited uptake is the \textbf{lack of practical assumptions}. Most of these approaches assume minimal structure on the returns (often treating them as \emph{adversarially} generated sequences) leading to strategies that have general worst-case theoretical guarantees but sometimes overly conservative for real-world scenarios. This assumption of adversarial returns is far from what is commonly observed in practice, where returns often exhibit patterns, correlations, and statistical properties that can be exploited for better performance.

Additionally, the \textbf{evaluation metric} used in OPS, namely the log-wealth or cumulative logarithmic return, is not commonly used by practitioners to assess portfolio performance. In practice, metrics such as risk-adjusted returns (\emph{e.g.} Sharpe ratio \citep{sharpe1994sharpe} or expected exponential utilities) are more commonly employed to assess and compare portfolio strategies. While the use of log-wealth is theoretically motivated by its asymptotic properties (e.g., maximizing the growth rate of wealth), its practical implications may not align well with the objectives of real-world investors. In this work, we aim to bridge this gap by incorporating more realistic assumptions about asset returns and developing a framework that is computationally feasible and practically aligned with investor objectives.

\section{Proofs of Section \ref{sec:theoretical_guarantees}}
\subsection{Auxiliary Lemmas}

\begin{lemma}\label{lemma:F_mf_closed}
    The mean-field space $\Fmf(\real^d\times\Theta)$ is closed in the space of probability measures $\cMplus$.
\end{lemma}
\begin{proof}
Consider any sequence $(\rho^i)_{i\in\mathbb{N}}$ in $\Fmf(\real^d\times\Theta)$ that has a limit in $\cMplus(\real^d\times\Theta)$. Then for any $i\in\mathbb{N}$, $\rho^i$ can be factorized as
\begin{align*}
    \rho^i(\dint(y, \theta)) = \rho_y^i(\dint y)\prod_{k=1}^K \rho_k^i(\dint\theta_k)\,,
\end{align*}
where we recall that we assume that $\Theta$ factorizes as a product of $K$ subspaces, $\Theta = \prod_{k=1}^K \Theta_k$. Then limit of $(\rho^i)_i$ exists by construction, and
\begin{align*}
    \lim_{i\to+\infty}\rho^i(\dint(y, \theta)) &= \lim_{i\to+\infty}\left(\rho_y^i(\dint y)\prod_{k=1}^K \rho_k^i(\dint\theta_k)\right)= \lim_{i\to+\infty}\rho_y^i(\dint y) \prod_{k=1}^K \lim_{i\to+\infty}\rho_k^i(\dint\theta_k)\in \Fmf(\real\times\Theta)\,,
\end{align*}
which proves that the limit of any convergent sequence in $\Fmf(\real^d\times\Theta)$ has its limit in $\Fmf(\real^d\times\Theta)$, and hence $\Fmf(\real\times\Theta)$ is closed in $\cMplus$.    
\end{proof}
    
\subsection{Proof of  \texorpdfstring{\cref{prop:convexity_of_psi}}{Lg}}\label{subsec:app_proof_convecity}

For any $\delta\in\cD$ and any measurable $h$ in $\real^d \times \Theta$, $h\mapsto g(h) = \sup_{\rho\in\cMplus(\cY\times\Theta)}(\langle h, \rho\rangle - \KL(\rho, \pi))$ is a convex map since it is defined as the Fenchel-Legendre transformation of $\rho\mapsto \KL(\rho, \pi_n)$, with $\rho\in\cMplus(\cY\times\Theta)$ (and the space of probability measures $\cMplus(\cY\times\Theta)$ is a convex set). Thus, we also have that the map $h\mapsto \Tilde{g}(h) = \sup_{\rho\in\Fmf(\cY\times\Theta)}(\langle h, \rho\rangle - \KL(\rho, \pi))$ is convex since for any probability measure $\rho\in\cMplus(\cY\times\Theta)$, $h\mapsto \langle h, \rho\rangle - \KL(\rho, \pi)$ is convex (as a linear map) and $\Tilde{g}$ is the \emph{pointwise supremum} of a family of convex function (the supremum conserves convexity \citep{boyd2004convex}). Taking h as $h_\delta(y) = -\lambda \delta^\top y$ and remarking that it is a convex function with respect to $\delta$ for a given $\lambda >0$ shows that $\obj$ is convex: indeed, by composition of convex functions, $\Tilde{g}(h_\delta(y))$ is convex with respect to $\delta$, and so $\obj(\delta) = \Tilde{g}(h_\delta(y))$. Moreover,  $\rho\mapsto \KL(\rho, \pi_n)$ is \emph{strongly convex} on the space of probability measures $\cMplus(\real^d\times\Theta)$, and hence $g$ is \emph{smooth} (as a convex conjugate of a strongly convex function). Hence, $\obj$ is also smooth by composition (with the same arguments as above for the convexity). 

The convergence rate mentioned follows directly from the classical results of gradient descent applied to convex, smooth functions (see, for instance, \citet[Chapter 5]{bach2024learning}).

\subsection{Proof of  \texorpdfstring{\cref{th:consistency}}{Lg}}\label{subsec:app_proof_consistency}
The first step involves interchanging the limit and the argmin over $\delta \in \mathcal{D}$, allowing us to express it as:
\begin{align*}
    \lim_{n\to+\infty}\argmin_{\delta\in\cD}\obj(\delta) \triangleq \argmin_{\delta\in\cD}\lim_{n\to+\infty}\obj(\delta)\,.
\end{align*}
This step $\triangle$ is non-trivial because the argmin function is a \textit{set}, necessitating the use of general regularity conditions. Additionally, $\obj$ is defined implicitly as a supremum over a space of measures. The second step involves analyzing $\lim_{n \to +\infty} \obj(\delta)$, for which we know how to proceed based on the statistical model introduced in \cref{sec:algorithm}.

\subsubsection{Inverting Limit and Argmin}
We rely on \citet[Theorem 7.33]{rockafellar2009variational} for this purpose; this strong result requires to show the two following conditions:
\begin{enumerate}
\item[\textcolor{blue}{\textbf{C.1.}}] $(\obj)_n$ epi-converges to $\obj^*$, where $\obj^*$ is lower-semi-continuous and proper.
\item[\textcolor{blue}{\textbf{C.2.}}] $(\obj)_n$ is a lower-semi-continuous and proper sequence.
\end{enumerate}
To show that $(\obj)_n$ epi-converges, we first establish some general regularity properties of this sequence, from which the epi-convergence will naturally follow.
For any $\lambda > 0$ and $n\in\mathbb{N}$, we introduce the functional
\begin{align}\label{eq:def_f_n}
    f_n(\delta, \rho) = -\lambda \delta^\top \E{\rho}{\Ypred} - \KL(\rho, \pi_n)\,.
\end{align}

\textbf{\bm{$(\obj)_n$} is a uniformly continuous sequence.} 
Since $\Theta$ is a Polish space, $\mathbb{R}^d \times \Theta$ is also a Polish space. According to \citet[Th. 1.3]{billingsley2013convergence}, $\cMplus(\mathbb{R}^d \times \Theta)$ is a space of \emph{tight} measures. By Prokhorov's theorem, $\cMplus(\mathbb{R}^d \times \Theta)$ is \emph{relatively compact} in the weak-* topology. Given that $\Fmf(\mathbb{R}^d \times \Theta)$ is closed in $\Fmf$ (\cref{lemma:F_mf_closed}), it follows that $\Fmf(\mathbb{R}^d \times \Theta)$ is also compact as a closed subset of a relatively compact space. By the \emph{Maximum theorem}, $\obj$ is continuous on $\cD$. Furthermore, for any $\rho \in \Fmf$, the function $f_n(\delta, \rho)$, as defined in \eqref{eq:def_f_n}, is linear with respect to $\delta$ (since the KL term is independent of $\delta$), making $f_n$ \emph{uniformly continuous} in $\delta$. Consequently, since $\obj$ is continuous on a compact set and $f_n$ is uniformly continuous with respect to $\delta$, we have that for all $\rho \in \Fmf$ and for any $\varepsilon > 0$, there exists a $\gamma > 0$ such that
\begin{align*}
    \lVert\delta_1 - \delta_2\rVert < \gamma \implies |f_n(\delta_1, \rho) -f_n(\delta_2, \rho)| \leq \epsilon\,,
\end{align*}
and therefore, 
\begin{align*}
\forall \varepsilon > 0,\, \exists\gamma >0,\, \lVert\delta_1 - \delta_2\rVert < \gamma\implies |\obj(\delta_1) - \obj(\delta_2)|\leq \sup_{\rho\in\Fmf}|f_n(\delta_1, \rho) -f_n(\delta_2, \rho)| \leq \epsilon
\end{align*}
independently in $n$, which is the definition of uniform continuity of $\obj$.

\textbf{\bm{$(\obj)_n$} is epi-convergent sequence.} 
Since $\obj$ is uniformly continuous, smooth, and convex on a compact domain for any $n \in \mathbb{N}$ (\cref{prop:convexity_of_psi}), it follows that $\obj$ is uniformly bounded on this compact space, implying that $(\obj)_n$ is \emph{equicontinuous}. Additionally, $(\obj)_n$ converges pointwise to a limit $\obj^*$, as established by \cref{assumption:contractance}. By the Arzelà-Ascoli theorem, the equicontinuous sequence $(\obj)_n$ converges uniformly to $\obj^*$. According to \citet[Theorem 7.11]{rockafellar2009variational}, $(\obj)_n$ epi-converges if and only if it converges continuously. Since uniform convergence implies continuous convergence, we conclude that $(\obj)_n$ is \emph{epi-convergent} to $\obj^*$, thereby verifying \textcolor{blue}{\textbf{C.1}}.

\textbf{Lower semi-continuity and proper conditions.}
Since $(\obj)_n$ is continuous, it is also lower semi-continuous. Furthermore, because $(\obj)_n$ converges continuously, $\obj^*$ is continuous and thus lower semi-continuous as well. To show that any preimage of a set $I \subset \mathbb{R}$ (e.g., a closed interval) is compact, note that since $\obj$ is continuous, $\obj^{-1}(I)$ is a closed subset of $\cD$. Given that $\cD$ is compact, $\obj^{-1}(I)$ is a closed subset of a compact space, and hence compact. The same reasoning applies to $\obj^*$ due to continuous convergence. Therefore, we have verified \textcolor{blue}{\textbf{C.2}}.

\subsubsection{Asymptotic Objective Function}
We now turn our attention to computing $\lim_{n \to +\infty} \obj = \obj^*$. Since $\rhoVB$ converges to a distribution $\hat{\rho}_\infty$, where $\hat{\rho}\infty$ is the fixed point of $T_\infty$ (\cref{assumption:contractance}), we can explicitly compute $\obj^*$ using the known form of $\obj$ as a function of $n$ for the statistical models under consideration. The following lemma formalizes this result.
\begin{lemma}\label{lemma:asymptotic objective_appendix}
For both \emph{GW} \eqref{eq:model_GW} and \emph{AR} \eqref{eq:model_AR1} models, we have
\begin{align*}
\forall \delta\in\cD\,,\quad\obj^*(\delta) = \frac{1}{2}(\lambda\delta)^\top \hat\Sigma^{-1}_\infty(\lambda\delta) - \lambda\delta^\top\hat\mu_\infty\,,
\end{align*}
where for the GW model we have
\begin{align*}
    &\hat\mu_\infty = \lim_{n\to+\infty}\left(\frac{1}{n}\sum_{t=1}^n Y_t\right)\,, &&\hat\Sigma_\infty =\lim_{n\to+\infty}\left(\frac{1}{n}\sum_{t=1}^n \left(Y_t - \hat\mu_\infty\right)\left(Y_t - \hat\mu_\infty\right)^\top\right)\,,
\end{align*}
and for the AR model
\begin{align*}
    &\hat\mu_\infty = \lim_{n\to+\infty}\left(\sum_{t=1}^n Y_t Y_{t-1}\right)\left(\sum_{t=1}^n Y_t Y_{t}\right)^{-1} Y_\infty\,,&&\hat\Sigma_\infty =\lim_{n\to+\infty}\left(\frac{1}{n}\sum_{t=1}^n \left(Y_t - \hat\mu_\infty\right)\left(Y_t - \hat\mu_\infty\right)^\top\right)\,,
\end{align*}
which is the corresponding sample mean estimate for the GW and the AR model respectively, and the corresponding sample covariance estimate.
Hence, the asymptotic variational decision writes
\begin{align*}
    \lim_{n\to+\infty}\deltaVB = \argmin_{\delta\in\cD}\left\{\frac{1}{2}(\lambda\delta)^\top \hat\Sigma^{-1}_\infty(\lambda\delta) - \lambda\delta^\top\hat\mu_\infty\right\}\,,
\end{align*}
i.e. the Markowitz decision in $\cD$.
\end{lemma}

We now prove \cref{lemma:asymptotic objective_appendix} for both GW and AR model.

\begin{proof}[Proof of \cref{lemma:asymptotic objective_appendix} for GW model]
Starting from \cref{lemma:rho_GW}, the asymptotic operator $T_\infty$ is the limit of the operator $T_n$ defined in \cref{lemma:rho_GW} when $n\to+\infty$ (by \cref{assumption:contractance}), giving 
\begin{align*}
T_\infty : (\xi_y, \Lambda_y, \xi_\mu, \Lambda_\mu, \psi_\Lambda) \mapsto \begin{pmatrix}
    \xi_\mu - \lambda(\nu\psi_\Lambda)^{-1}\delta\\
    \nu\psi_\Lambda\\
    \lim_{n\to+\infty}\left(\frac{1}{n}\sum_{t=1}^n Y_t\right)\\
    n\nu\psi_\Lambda\\
    \lim_{n\to+\infty}\left( n\left(\Lambda_\mu^{-1} +\xi_\mu\xi_\mu^\top\right) + \sum_{t=1}^n Y_t Y_t^\top - 2\sum_{t=1}^n Y_t\xi_\mu^\top \right)^{-1}
\end{pmatrix}\,.
\end{align*}
Thanks to \cref{assumption:contractance}, the fixed point of $T_\infty$ denoted by $(\xi^\infty_y, \Lambda^\infty_y, \xi_\mu, \Lambda^\infty_\mu, \psi^\infty_\Lambda)$ satisfies
\begin{align*}
\begin{cases} 
\xi^\infty_y = \hat\mu_\infty - \lambda \hat\Sigma_\infty \delta\\
\Lambda_y^\infty = \hat\Lambda_\infty\\
\xi_\mu^\infty = \hat\mu_\infty\\
\Lambda_\mu^\infty = \lim_{n\to+\infty}n\hat\Lambda_\infty\\
\psi^\infty_\Lambda = \left(\sum_{s=1}^n (Y_t - \hat\mu_\infty)(Y_t - \hat\mu_\infty)^\top\right)^{-1}\,,
\end{cases}
\end{align*}
where
\begin{align*}
    &\hat\mu_\infty = \lim_{n\to+\infty}\left(\frac{1}{n}\sum_{t=1}^n Y_t\right)
    &&\hat\Sigma_\infty =\lim_{n\to+\infty}\left(\frac{1}{n}\sum_{t=1}^n \left(Y_t - \hat\mu_\infty\right)\left(Y_t - \hat\mu_\infty\right)^\top\right)\,,&& \hat\Lambda_\infty = \hat\Sigma_\infty^{-1}  \,. 
\end{align*}

Plugging this solution the objective function in \cref{lemma:objective_function_standard_GW} and keeping only terms depending on $\delta$, the asymptotic objective function writes
\begin{align*}
\obj^*(\delta) &= -\lambda\delta^\top\left(\hat\mu_\infty - \lambda \hat\Sigma_\infty\delta\right)\,.
\end{align*}
\end{proof} 

\begin{proof}[Proof of \cref{lemma:asymptotic objective_appendix} for AR model]
Starting from \cref{lemma:rho_Ar}, the asymptotic operator $T_\infty$ writes
\begin{align*}
    T_\infty:\left(\xi_y, \Lambda_y, M_\Gamma, V_\Gamma \otimes U_\Gamma, \psi_\Lambda^{-1}\right) \mapsto\begin{pmatrix}
        M_\Gamma Y_\infty - \lambda(\nu\psi_\Lambda)^{-1}\delta\\
        \nu_\Lambda \psi_\Lambda\\
        \lim_{n\to+\infty}\left(\sum_{t=1}^n Y_t Y_{t-1}\right)\left(\sum_{t=1}^n Y_t Y_{t}\right)^{-1} \\
        \lim_{n\to+\infty}\left(\sum_{t=1}^n Y_t Y_t^\top \otimes \nu_\Lambda \psi_\Lambda\right)^{-1}\\
        \lim_{n\to+\infty} \left(\sum_{t=0}^n Y_t Y_t^\top - 2\left(\sum_{t=1}^n Y_{t-1}Y_t^\top\right)^\top \left(\sum_{t=1}^n Y_t Y_t^\top\right)^{-1}\left(\sum_{t=1}^n Y_{t-1}Y_t^\top\right)\right)
    \end{pmatrix}\,,
\end{align*}
where $Y_\infty$ denotes the last observation of the dataset.

Thanks to \cref{assumption:contractance}, the fixed point of $T_\infty$ denoted by $\left(\xi_\infty, \Lambda_\infty, M^\infty_\Gamma, (V_\Gamma \otimes U_\Gamma)^\infty, \psi_\Lambda^{\infty}\right)$ satisfy
\begin{align*}
\begin{cases} 
\xi^\infty_y = \hat\mu - \lambda \hat\Sigma_\infty \delta \\
\Lambda_y^\infty = \hat\Lambda_\infty\\
M_\Gamma^\infty = \lim_{n\to+\infty}\left(\sum_{t=1}^n Y_t Y_{t-1}\right)\left(\sum_{t=1}^n Y_t Y_{t}\right)^{-1}  \\
(V_\Gamma \otimes U_\Gamma)^\infty = \lim_{n\to+\infty}\left(\sum_{t=1}^n Y_t Y_t^\top \otimes \left(\sum_{t=0}^n Y_t Y_t^\top - 2\frac{1}{n}\left(\sum_{t=1}^n Y_{t-1}Y_t^\top\right)^\top \left(\sum_{t=1}^n Y_t Y_t^\top\right)\left(\sum_{t=1}^n Y_{t-1}Y_t^\top\right)\right)^{-1} \right)^{-1} \\
\psi^\infty_\Lambda = \lim_{n\to+\infty} \left(\frac{1}{n}\left(\sum_{t=0}^n Y_t Y_t^\top - 2\frac{1}{n}\left(\sum_{t=1}^n Y_{t-1}Y_t^\top\right)^\top \left(\sum_{t=1}^n Y_t Y_t^\top\right)\left(\sum_{t=1}^n Y_{t-1}Y_t^\top\right)\right)^{-1}\right)  \,,
\end{cases}
\end{align*}
where
\begin{align*}
    &\hat\mu_\infty = \lim_{n\to+\infty}\left(\sum_{t=1}^n Y_t Y_{t-1}\right)\left(\sum_{t=1}^n Y_t Y_{t}\right)^{-1} Y_\infty\,,&&\hat\Sigma_\infty = \sum_{t=0}^n Y_t Y_t^\top - 2\frac{1}{n}\left(\sum_{t=1}^n Y_{t-1}Y_t^\top\right)^\top \left(\sum_{t=1}^n Y_t Y_t^\top\right)\left(\sum_{t=1}^n Y_{t-1}Y_t^\top\right) \\
    &\hat\Lambda_\infty = (\hat\Sigma_\infty)^{-1}\,.
\end{align*}
Plugging this solution to the objective function (\cref{lemma:objective_function_AR1}) and keeping only terms depending on $\delta$, the asymptotic objective function writes
\begin{align*}
\lim_{n\to+\infty}\obj^*(\delta) &= -\lambda\delta^\top\left(\hat\mu_\infty - \lambda \hat\Sigma_\infty\delta\right)\,.
\end{align*}
\end{proof}

\section{Derivation of \texorpdfstring{$\algVB$}{Lg} for Specific Models: Fixed-Point Operators and Objective Functions}\label{sec:alg_specific}\label{sec:app_complete_statements}
Let us begin by introducing some convenient compact notations.
\begin{definition}[Kronecker product]
Let $A\in\real^{n\times p}$ and $B\in\real^{m\times q}$. Then the \emph{kronecker product} between $A$ and $B$, denoted by $A\otimes B\in\real^{mn\times pq}$ is defined as follows,
\begin{align*}
    A\otimes B = \begin{pmatrix}
        a_{1, 1}B & \hdots & a_{1, p}B\\
        a_{2, 1}B & \hdots & a_{2, p}B\\
        \vdots & \vdots & \vdots\\
        a_{n, 1}B & \hdots & a_{n, p}B
    \end{pmatrix}\,.
\end{align*}
\end{definition}

\begin{definition}[Vectorization operator]
Let $A\in\real^{n\times p}$, such that
\begin{align*}
    A = \begin{pmatrix}
        a_{1, 1}&\hdots & a_{1, p}\\
        \vdots & \vdots & \vdots\\
        a_{n, 1} & \hdots & a_{n, p}
    \end{pmatrix} = \begin{pmatrix}
        \bm{a_1} & \hdots & \bm{a_p}
    \end{pmatrix}\,.
\end{align*}
Then, we denote $\ve(A)$ the vector of size $np$ such that $ \ve(A) = \begin{pmatrix}
        \bm{a_{1}}\\
        \vdots \\
        \bm{a_{p}}
    \end{pmatrix}$.
\end{definition}

\begin{remark}
Using \cref{def:matrix normal distribution}, the \emph{AR} model \eqref{eq:model_AR1} is \emph{equivalent} to the following formulation,
\begin{align*}
    &Y_t \condi Y_{t-1}, \Gamma, \Lambda \sim \cN(\Gamma Y_{t-1}, \Lambda) \qquad \forall t \in \mathbb{N}^*\\
    &\ve(\Gamma) \sim \cN(\ve(M_0), V_0\otimes U_0)\,,\quad\Lambda\sim\cW(\nu_0, \psi_0)\,.
\end{align*}    
\end{remark}

\begin{proposition}[Properties of $\ve$ operator]
\label{prop:vectorization_properties}\textcolor{white}{xx}
    \begin{itemize}
        \item $\ve(AXB) = (B^\top \otimes A)\ve(X)$.
        \item $\Tra(ABC) = \ve(A^\top)^\top (I\otimes B)\ve(C)$.
        \item $\Tra(A^\top BCD^\top) = \ve(A)^\top (D\otimes B)\ve(C)$.
    \end{itemize}
\end{proposition}
We refer to \citet{quintana1987multivariate} for the proofs of these  results.

\subsection{General Fixed-Point Equation (Proof of  \texorpdfstring{\cref{prop:fixed_point}}{Lg})}
This proof is a direct application of \citet[Chapter 10;][]{bishop2006pattern} to our problem; we have an additional risk term $e^{-\lambda \delta^\top y}$, which modifies the computation of the complete joint distribution. We recall that $\Tilde{\pi}_n$ is the probability distribution defined as $\dint\Tilde{\pi}_n = \frac{e^{-\lambda \delta^\top \Ypred}}{\E{\pi_n}{e^{-\lambda \delta^\top \Ypred}}}\dint\pi_n$. Then for any $\rho\in\cMplus$, we have
\begin{align*}
    \KL(\rho, \Tilde{\pi}_n) = \log\pi(H_n) - \mathrm{E}(\rho)+\log Z_\delta\,, 
\end{align*}
where
\begin{align*}
    \mathrm{E}(\rho) = \int_{\cY\times\Theta} \log\left(\frac{e^{-\lambda \delta^\top \Ypred}\pi\left(\Ypred, \theta, H_n\right)}{\rho(\Ypred, \theta)}\right)\rho(\dint(\Ypred,\theta))\,.
\end{align*}
$E(\rho)$ is called the \textit{evidence lower bound} (ELBO), and is the only term that depends on $\rho$. Hence, minimizing $\rho \mapsto \KL(\rho, \Tilde{\pi}_n)$ is equivalent at maximizing $\rho\mapsto\rm{E}(\rho)$. Then, for any $\rho\in\Fmf$, we have
\begin{align}\label{eq:temp_elbo}
    \rm{E}(\rho) = \int_{\cY\times\Theta}\left(\log\left(e^{-\lambda \delta^\top \Ypred}\pi\left(\Ypred, \theta, H_n\right)\right) - \log\rho_y(\Ypred) - \sum_{i=1}^K \rho_i(\theta_i)\right)\rho_y(\dint\Ypred)\prod_{i=1}^K \rho_i(\dint\theta_i)\,.
\end{align}
Keeping terms that depend on $\theta_j$ only and applying Fubini theorem, we have
\begin{align*}
    \rm{E}(\rho) &\propto_{\theta_j} \int_{\Theta_j}\left(\int_{\cY\times\Theta\setminus\Theta_j}e^{-\lambda \delta^\top \Ypred}\pi\left(\Ypred, \theta, H_n\right)\rho_y(\dint\Ypred)\prod_{i\neq j}^K \rho_i(\dint\theta_i)\right)\rho_j(\dint\theta_j) - \int_{\Theta_j}\log\rho_j(\dint\theta_j)\rho_j(\dint{\theta_j})\,.
\end{align*}
The maximizer of $\rho\mapsto \rm{E}(\rho)$ with respect to each of the $\theta_i$'s can be derived (by computing the Lagragian of \eqref{eq:temp_elbo} \citep{jordan1999introduction}), and we can show that the maximum is reached when
\begin{align}\label{eq:elbo_solution}
    \forall j\in[K]\,,\quad \log\rho_j(\dint\theta_j)\propto \exp\left(\int_{\cY\times\Theta\setminus\Theta_j}e^{-\lambda \delta^\top \Ypred}\pi\left(\Ypred, \theta, H_n\right)\rho_y(\dint\Ypred)\prod_{i\neq j}^K \rho_i(\dint\theta_i)\right)\,,
\end{align}
which can be seen as the expectation of $e^{-\lambda \delta^\top \Ypred}\pi\left(\Ypred, \theta, H_n\right)$ taken with respect to all parameters $\theta_i$ with measure $\rho_i$ except $\theta_j$. The main advantage of using mean-field assumption is that \eqref{eq:elbo_solution} yields to a natural algorithm where we update successively each $\rho_i$'s until stabilization.

\subsection{The Gaussian-Gaussian Model}\label{subsec:app_Gaussian_Gaussian}
We begin with a straightforward example where the covariance matrix $\Sigma_*$ is assumed to be \emph{deterministically} known. This setting is not realistic since estimating $\Sigma_*$ is one main goal of portfolio selection. Given the conjugate nature of this model, we can directly compute the optimal Bayes decision $\delta^*$ by explicitly calculating and minimizing the risk function $\risk$. We show how to instantiate $\algVB$ for this model primarily for illustrative purposes.

By putting a prior distribution on the mean, the Gaussian-Gaussian model writes
\begin{align}\label{eq:model_Gaussian_Gaussian}
&Y_t \condi \mu \overset{\rm{i.i.d.}}{\sim} \cN(\mu, \Sigma_*)&&\forall t\in \mathbb{N}\nonumber \\
& \mu\sim\cN(\mu_0, \Lambda_0^{-1})\,.
\end{align}

\subsubsection{Computing Directly  \texorpdfstring{$\risk$}{Lg}} 
In this (conjugate) model, the posterior predictive $\pi(\dint\Ypred\condi H_n)$ and the posterior parameter $\pi(\dint\Ypred\condi H_n)$ can be directly computed explicitly,
\begin{align*}
    &\pi(\dint\mu\condi H_n) = \cN\left(\dint\theta ; \hat m_\mu, \hat\Sigma_\mu \right)\,; && \pi(\dint\Ypred\condi H_n) = \cN\left(\dint\theta ; \hat m_{\Ypred}, \hat\Sigma_{\Ypred} \right)\,,
\end{align*}
where (by completing the Gaussian square, \emph{e.g.} see \citet[Chapter 2;][]{bishop2006pattern})
\begin{align*}
    &\hat m_\mu = \left( \Sigma_*^{-1} + \frac{1}{n}\Sigma_0^{-1} \right)^{-1}\left( \Sigma_*^{-1}\hat\mu_n + \frac{1}{n}\Sigma_0^{-1}\mu_0 \right) &&\hat\Sigma_\mu^{-1} =\frac{1}{n}\left( \Sigma_*^{-1} + \frac{1}{n}\Sigma_0^{-1} \right)\\
    &m_{\Ypred} = \hat m_\mu && \hat\Sigma_{\Ypred} = \Sigma_* + \hat\Sigma_\mu\,.
\end{align*}
Therefore, in this model, the risk function $\risk$ given in \eqref{eq:optimal_Bayesian_decision_exp} can be directly computed,
\begin{align*}
    \risk(\delta) &= \int_{\real^d}e^{-\lambda \delta^\top \Ypred}\cN(\dint\Ypred ; m_{\Ypred}, \hat\Sigma_{\Ypred}) = C\exp\left( \delta^\top \hat\Sigma_{\Ypred}\delta - 2\hat m_\mu^\top \delta \right)\,,
\end{align*}
where $C$ is a constant that does not depend on $\delta$, and $\risk(\delta)$ is minimized when
\begin{align}\label{eq:bayes_decision_gaussian}
    \delta^* = \argmin_{\delta\in\cD}\left\{ -\lambda \delta^\top \hat\Sigma_{\Ypred}\delta + 2\hat m_\mu^\top \delta \right\}\,.
\end{align}

\subsubsection{ \texorpdfstring{$\algVB$}{Lg} Instantiated on Gaussian-Gaussian Model} 
While we can explicitly compute \eqref{eq:bayes_decision_gaussian}, we outline how $\algVB$ operates for this simple model to provide a clear illustration. The first goal is to compute the fixed-point operator for this model (defined implicitely in \cref{prop:fixed_point}). For this model, since the estimation only focuses on the mean return $\mu$, the mean-field family writes
\begin{align*}
    \Fmf(\real^d\times\real^d) = \left\{\rho \in \cMplus(\real^d\times \real^d )\;:\;\rho(\dint(\Ypred, \mu)) = \rho_y(\dint\Ypred)\rho_\mu(\dint\mu) \,,\,\rho_y,\rho_\mu\in\cMplus(\real^d) \right\}\,,
\end{align*}
and the joint posterior ${\pi}_n$ can be written as
\begin{align*}
    {\pi}_n(\dint(\Ypred, \mu)) = \pi(\dint(\Ypred, \mu)\condi H_n)\propto_{\Ypred, \mu} \pi(H_n\condi \mu)\pi(\dint\Ypred\condi \mu)\pi_0(\dint\mu)\,.
\end{align*}
Therefore, from \cref{prop:fixed_point}, the variational distribution for $\Ypred$ is given by
\begin{align*}
    \log\rho_y(\Ypred) &\propto_{\Ypred} -\lambda\delta^\top \Ypred -\frac{1}{2}\E{\rho_\mu}{\left(\Ypred - \mu \right)^\top \Sigma_*^{-1}\left(\Ypred - \mu \right)}\\
    &\propto_{\Ypred} -\lambda \delta^\top \Ypred - \frac{1}{2}\Ypred^\top \Sigma_*^{-1}\Ypred + \Ypred\Sigma_*^{-1}\E{\rho_\mu}{\mu}\\
    &\propto -\frac{1}{2}\left( \Ypred^\top \Sigma_*^{-1}\Ypred - 2\Ypred^\top\left( \Sigma_*^{-1}\E{\rho_\mu}{\mu} - \lambda \delta \right) \right)\\
    &\propto \cN\left( \Ypred ; \xi_y, \Lambda_y^{-1} \right)\,,
\end{align*}
where $\Lambda_y = \Sigma_*^{-1}$ and $\xi_y = \E{\rho_\mu}{\mu} - \lambda \Sigma_*\delta$, and by doing the same as above, the variational distribution for parameter $\mu$ is given by
\begin{align*}
    \log\rho_\mu(\mu) &\propto_{\mu} -\lambda \delta^\top \E{\rho_y}{\Ypred} -\frac{1}{2} \sum_{t=1}^n \left(Y_t - \mu\right)^\top \Sigma_*^{-1}\left(Y_t - \mu\right) -\frac{1}{2} \E{\rho_y}{\left(\Ypred - \mu\right)^\top \Sigma_*^{-1}\left(\Ypred - \mu\right)} \\
    &-\frac{1}{2} \left(\mu - \mu_0\right)^\top \Sigma_0^{-1} \left(\mu - \mu_0\right)\\
    &\propto_{\mu}-\frac{1}{2}\left( \sum_{t=1}^n \left(Y_t -\mu\right)^\top \Sigma_*^{-1}\left(Y_t -\mu\right) + \E{\rho_y}{\left(\Ypred - \mu\right)\Sigma_*^{-1}\left(\Ypred - \mu\right)} + \left(\mu - \mu_0\right)^\top \Sigma_0^{-1} \left(\mu - \mu_0\right)\right)\\
    &\propto -\frac{1}{2}\left( \mu^\top \left( (n+1)\Sigma_*^{-1} + \Sigma_0^{-1} \right)\mu - 2\mu^\top \left(\Sigma_*^{-1}\left(\sum_{t=1}^n Y_t + \E{\rho_y}{\Ypred}\right)+\Sigma_0^{-1}\mu_0\right) \right)\\
    &\propto \cN\left(\mu ; \xi_\mu, \Lambda_\mu^{-1}\right)\,,
\end{align*}
where
\begin{align}\label{eq:temp_Gaussian_Gaussian_2}
    (&\Lambda_\mu = (n+1)\Sigma_*^{-1} + \Sigma_0^{-1} && \xi_\mu = \Lambda_\mu^{-1}\left(\Sigma_*^{-1}\left(\sum_{t=1}^n Y_t + \E{\rho_y}{\Ypred}\right)+\Sigma_0^{-1}\mu_0\right)\,.
\end{align}
Combining these equations leads to the following fixed-point system $(T_n)$,
$$
(T_n) : \begin{cases}
\xi_y = \xi_\mu - \lambda \Sigma_*\delta\\
\Lambda_y = \Sigma_*^{-1}\\
\xi_\mu = \frac{1}{n+1}\left( \Sigma_*^{-1} + \frac{1}{n+1}\Sigma_0^{-1} \right)^{-1}\left( \Sigma_*^{-1}\sum_{t=1}^n Y_t + \Sigma_0^{-1}\mu_0 \right) + \frac{1}{n+1}\left(\Sigma_*^{-1}+\frac{1}{n+1}\Sigma_0^{-1}\right)^{-1}\Sigma_*^{-1}\xi_y  \\
\Lambda_\mu = (n+1)\Sigma_*^{-1} + \Sigma_0^{-1}
\end{cases} \,.
$$
The remarkable property of the Gaussian-Gaussian case is that the operator $T_n$ has an unique fixed-point; defining
\begin{align}\label{eq:temp_gaussian_gaussian}
&\alpha = \frac{1}{n+1}\left( \Sigma_*^{-1} + \frac{1}{n+1}\Sigma_0^{-1} \right)^{-1}\left( \Sigma_*^{-1}\sum_{t=1}^n Y_t + \Sigma_0^{-1}\mu_0 \right) \nonumber\\
&C =  \frac{1}{n+1}\left(\Sigma_*^{-1}+\frac{1}{n+1}\Sigma_0^{-1}\right)^{-1}\Sigma_*^{-1}\,,
\end{align}
we have
$$
\begin{cases}
\xi_y = \xi_\mu - \lambda \Sigma_*\delta\\
\xi_\mu = \alpha + C\xi_y
\end{cases} 
\iff 
\begin{cases}
\xi_y = \left(I_d-C\right)^{-1}\alpha - \left(I_d-C\right)^{-1}\lambda \Sigma_*\delta\\
\xi_\mu = \alpha + \left(C^{-1}- I_d\right)^{-1}\alpha - \left(C^{-1} - I_d\right)^{-1}\lambda\Sigma_*\delta
\end{cases} \,.
$$

The second step is to compute the objective function based on the parameters $(\xi_y, \Lambda_y, \xi_\mu, \Lambda_mu)$ defined by $T_n$ by plugging these parameters to \eqref{eq:def_delta_vb}; for any $\delta\in\cD$,
\begin{align*}
    \obj(\delta) &= -\int_{\real^d\times\real^d} \log\left(\frac{\rho_y(\Ypred)\rho_\mu(\mu)}{\frac{e^{-\lambda \delta^\top \Ypred}\pi(\dint(\Ypred, \mu\condi H_n))}{\E{\pi_n}{e^{-\lambda\delta^\top}}}}\right) - \log\E{\pi_n}{e^{-\lambda\delta^\top}}\\
    &\propto_\delta -\E{\rho_y}{\log\rho_y(\Ypred)} - \E{\rho_\mu}{\log\rho_\mu(\mu)} - \lambda \delta^\top\E{\rho_y}{\Ypred} + \E{\rho_y, \rho_\mu}{\log\pi_n(\Ypred, \mu)}\\
    &\propto -\frac{1}{2}\log|\Lambda_y| - \frac{1}{2}\log|\Lambda_\mu| -\lambda \delta^\top \xi_y +\E{\rho_y, \rho_\mu}{\log\pi_n(\Ypred, \mu)}\,,
\end{align*}
with
\begin{align*}
    \E{\rho_y, \rho_\mu}{\log\pi_n(\Ypred, \mu)} \propto_\delta -\frac{1}{2}\xi_\mu^\top \Sigma_0^{-1}\xi_\mu + \xi_\mu^\top \Sigma_0^{-1}\mu_0\,.
\end{align*}
Since $\Lambda_y$ and $\Lambda_\mu$ do not depend on $\delta$ and given the expressions of the parameters $\xi_y$ and $\xi_\mu$ defined above, we have 
\begin{align}\label{eq:temp_solving_in_rd}
    \risk(\delta)&\propto_\delta - \lambda \delta^\top \xi_y - \frac{1}{2}\xi_\mu^\top \Sigma_0^{-1}\xi_\mu \,,
\end{align}
and after a few simplifications, we obtain
\begin{align*}
\risk(\delta)\propto_\delta (\lambda \delta)^\top \left( \Sigma_* - \frac{1}{2}\Sigma_*\left(I_d - C\right)^{-1} \right)(\lambda \delta) + (\lambda \delta)^\top \left( (C - I_d)^{-1}\alpha \right) \,.
\end{align*}

\textbf{Asymptotic analysis when $\cD = \real^d$.} From \eqref{eq:temp_gaussian_gaussian}, we have $\lim_{n\to+\infty}C = 0$ and $\lim_{n\to+\infty}\alpha = \hat\mu_{\infty}$, where $\mu_\infty = \lim_{n\to+\infty}\frac{1}{n}\sum_{t=1}^n Y_t$. Since \eqref{eq:temp_solving_in_rd} is a convex optimization problem when $\cD = \real^d$, we have 
\begin{align*}
    \lim_{n\to+\infty}\obj(\delta) = \frac{1}{2}(\lambda \delta)^\top \Sigma_*(\lambda \delta) - \lambda \delta^\top \hat\mu_\infty\,,
\end{align*}
which admits the unique minimizer
\begin{align*}
    \lim_{n\to+\infty}\deltaVB = \frac{1}{\lambda}\Sigma_*^{-1}\hat\mu_\infty\,,
\end{align*}
which is the same decision than the asymptotic Optimal Bayes decision \eqref{eq:bayes_decision_gaussian} on $\cD = \real^d$,
\begin{align*}
    \lim_{n\to+\infty}\delta^* = \frac{1}{\lambda}\Sigma_*^{-1}\hat\mu_\infty = \lim_{n\to+\infty}\deltaVB \,.
\end{align*}

\subsection{ \texorpdfstring{$\algVB$}{Lg} for the Gaussian-Wishart Model}\label{Appendix_Gaussian_Wishart}
We start by deriving the fixed-point equation, and then we derive the corresponding objective function $\obj$.
\begin{lemma}[Solution of \eqref{eq:def_rho_vb} under Gaussian-Wishart model]\label{lemma:rho_GW}
Under the stationary Gaussian-Wishart model \eqref{eq:model_GW}, for any $\delta\in\cD$, the corresponding variational distribution $\rhoVB$ can be factorised as follows,
\begin{align*}
    \rhoVB(\dint(\Ypred, \mu, \Lambda)) =  \rho_y(\dint\Ypred)\rho_\mu(\dint\mu)\rho_\Lambda(\dint\Lambda)\,,
\end{align*}
where $\rho_y(\dint\Ypred) = \cN(\dint\Ypred;\xi_y, \Lambda_y^{-1})$, $\rho_\mu(\dint\mu) = \cN(\dint\mu;\xi_\mu, \Lambda_\mu^{-1})$ and $\rho_\Lambda(\dint\Lambda) = \cW(\dint \Lambda ; \nu_\Lambda, \psi_\Lambda)$. Moreover, 
the variational parameters $(\xi_y, \Lambda_y,\xi_\mu, \Lambda_\mu, \nu_\Lambda, \psi_\Lambda)$ satisfy a fixed-point equation $T_n(\xi_y, \Lambda_y,\xi_\mu, \Lambda_\mu, \nu_\Lambda, \psi_\Lambda) = (\xi_y, \Lambda_y,\xi_\mu, \Lambda_\mu, \nu_\Lambda, \psi_\Lambda)$, where $T_n$ is given as follows:
\begin{align*}
T_n: (\xi_y, \Lambda_y,\xi_\mu, \Lambda_\mu, \nu_\Lambda, \psi_\Lambda) \mapsto \begin{pmatrix}
    \xi_\mu - \frac{\lambda}{\nu_\Lambda}\psi_\Lambda^{-1}\delta  \\
    \nu_\Lambda\psi_\Lambda\\
    \frac{1}{n+1}\big( \nu_\Lambda\psi_\Lambda + \frac{1}{n+1}\Lambda_0 \big)^{-1}\left( \nu_\Lambda\psi_\Lambda \big( \xi_y + \sum_{t\in[n]}Y_t \big)+\Lambda_0\mu_0 \right)\\
    (n+1)\nu_\Lambda\psi_\Lambda + \Lambda_0\\
    n+\nu_0+1\\
    \left( \Lambda_y^{-1} + \xi_y \xi_y^\top + (n+1)(\Lambda_\mu^{-1} + \xi_\mu\xi_\mu^\top) + \sum_{t\in[n]}Y_tY_t^\top - 2(\xi_y + \sum_{t\in[n]}Y_t)\xi_\mu^\top) + \psi_0^{-1} \right)^{-1}
\end{pmatrix}\,.
\end{align*}
\end{lemma}

\begin{proof}
We first express $\Tilde{\pi}_n$ independently of the underlying statistical model;
\begin{align}\label{eq:GW_pi_n_temp}
    \Tilde{\pi}_n(\Ypred, \mu, \Lambda) &\propto e^{-\lambda \delta^\top \Ypred}\pi(\Ypred, \mu, \Lambda \condi H_n)\nonumber\\
    &\propto e^{-\lambda \delta^\top \Ypred}\pi(H_n\condi Y_{n+1}, \mu, \Lambda)\pi(\Ypred, \mu, \Lambda)\nonumber\\
     &\propto e^{-\lambda \delta^\top \Ypred}\pi(H_n\condi Y_{n+1}, \mu, \Lambda)\pi(\Ypred\condi \mu, \Lambda)\pi_0(\mu, \Lambda)\,,
\end{align}
where the first equation follows from the definition of $\Tilde{\pi}_n$ in \cref{th:main_th} and the second equation follows from Bayes rule. Since the model \eqref{eq:model_GW} involves $n$ \emph{i.i.d.} observations conditionally on the parameters, \eqref{eq:GW_pi_n_temp} gives
\begin{align*}
    \Tilde{\pi}_n(\Ypred, \mu, \Lambda) &\propto e^{-\lambda \delta^\top \Ypred}\pi(H_n\condi Y_{n+1}, \mu, \Lambda)\pi(\Ypred\condi \mu, \Lambda)\pi_0(\mu, \Lambda)\\
    & \propto e^{-\lambda \delta^\top \Ypred} \left(\prod_{t=1}^n \cN(Y_t ;\mu, \Lambda^{-1})\right) \cN(\Ypred;\mu, \Lambda^{-1})\cN(\mu;\mu_0, \Lambda_0^{-1})\cW(\Lambda ; \nu_0, \psi_0)\,.
\end{align*}
By \cref{prop:fixed_point}, we can compute each variational distribution $\rho_y$, $\rho_\mu$ and $\rho^*_\Lambda$:
\begin{align*}
    \log\rho_y(\Ypred) &\propto_{\Ypred} \E{\rho_\mu, \rho_\Lambda}{\log\Tilde{\pi}_n(\Ypred,\mu, \Lambda)}\\
    &\propto \E{\rho_\mu, \rho_\Lambda}{ -\frac{1}{2}\left( 2\lambda \delta^\top \Ypred + (\Ypred - \mu)^\top \Lambda (\Ypred - \mu) \right) }\\
    &\propto -\frac{1}{2}\left( \Ypred^\top \E{\rho_\Lambda}{\Lambda}\Ypred - 2\Ypred^\top \left(\E{\rho_\Lambda}{\Lambda}\E{\rho_\mu}{\mu}-\lambda \delta\right) \right)\,,
\end{align*}
which gives, by completing the Gaussian square with respect to the variable $\Ypred$,
\begin{align}\label{eq:stationary_GW_rho_y}
    \rho_{\Ypred}(\dint\Ypred) &= \cN(\Ypred ; \xi_y, \Lambda_y^{-1})\\
    \text{where }&\xi_y = \E{\rho_\mu}{\mu} - \lambda\left(\E{\rho_\Lambda}{\Lambda}\right)^{-1}\delta \nonumber\\
    &\Lambda_y = \E{\rho_\Lambda}{\Lambda}\,.\nonumber
\end{align}
For the mean variational distribution $\rho_\mu$,
\begin{align*}
    \log\rho_\mu(\mu) &\propto_{\mu} \E{\rho_y, \rho_\Lambda}{\log\left(\prod_{t=1}^n \cN(Y_t; \mu, \Lambda^{-1})\cN(\Ypred;\mu, \Lambda^{-1})\cN(\mu;\mu_0, \Lambda_0^{-1}) \right)}\\
    &\propto -\frac{1}{2}\left( \mu^\top \E{\rho_\Lambda}{\Lambda}\mu - 2\mu^\top\E{\rho_{\Lambda}}{\Lambda}\E{\rho_{\Ypred}}{\Ypred}+ n \mu^\top \E{\rho_{\Lambda}}{\Lambda}\mu - 2\mu^\top \E{\rho_{\Lambda}}{\Lambda}\sum_{t=1}^n Y_t +\mu^\top \Lambda_0\mu - 2\mu^\top \Lambda_0\mu_0 \right)\\
    &\propto -\frac{1}{2}\left(\mu^\top \left(\E{\rho_{\Lambda}}{\Lambda}+n\E{\rho_{\Lambda}}{\Lambda}+\Lambda_0\right)\mu - 2\mu^\top\left(\E{\rho_{\Lambda}}{\Lambda}\E{\rho_{\Ypred}}{\Ypred}+\E{\rho_{\Lambda}}{\Lambda}\sum_{t=1}^n +\Lambda_0\mu_0\right)\right)\,,
\end{align*}
which gives, by completing the Gaussian square with respect to the variable $\mu$,
\begin{align*}
    \rho_\mu(\dint\mu) &= \cN(\dint\mu ; \xi_\mu, \Lambda_\mu^{-1})\\
    \text{where } &\xi_\mu = \left((n+1)\E{\rho_{\Lambda}}{\Lambda} + \Lambda_0\right)^{-1}\left( \E{\rho_{\Lambda}}{\Lambda}\left(\E{\rho_{\Ypred}}{\Ypred}+\sum_{t=1}^n\right)+\Lambda_0\mu_0 \right)\\
    &\Lambda_\mu = (n+1)\E{\rho_{\Lambda}}{\Lambda} + \Lambda_0\,.
\end{align*}
\end{proof}
Finally, for the precision variational distribution $\rho_\Lambda$,
\begin{align*}
    \log\rho_\Lambda(\Lambda)&\propto_{\Lambda} \E{\rho_{\Ypred}, \rho_{\mu}}{ \log\left(\prod_{t=1}^n \cN(Y_t ; \mu, \Lambda^{-1})\cN(\Ypred ; \mu, \Lambda^{-1})\cW(\Lambda ; \nu_0, \psi_0)\right) }\\
    &\propto \E{\rho_{\Ypred},\rho_{\mu}}{ -\frac{1}{2}\sum_{t=1}^n (Y_t - \mu)\Lambda(Y_t - \mu) + \frac{n}{2}\log|\Lambda| -\frac{1}{2}(\Ypred - \mu)^\top\Lambda(Y_t - \mu) }\\
    &+ \frac{1}{2}\log|\Lambda| + \frac{\nu_0 - d + 1}{2}\log|\Lambda| - \frac{1}{2}\Tra(\psi_0^{-1}\Lambda)\\
    &\propto \mathbb{E}_{\rho_{\Ypred}, \rho_\mu}\bigg[-\frac{1}{2}\left( \E{\rho_{\Ypred}}{\Tra(\Ypred\Ypred^\top\Lambda)} \right) -2\E{\rho_\mu}{\mu}^\top\Lambda \E{\rho_{\Ypred}}{\Ypred} + \E{\rho_{\Ypred}}{\Tra(\mu\mu^\top\Lambda)} \\
    &+\Tra(\sum_{t=1}^n Y_t Y_t^\top \Lambda) - 2\E{\rho_\mu}{\Tra(\mu\sum_{t=1}^n Y_t^\top \Lambda)} + \E{\rho_\mu}{n\Tra(\mu\mu^\top \Lambda)} +\Tra(\psi_0^{-1}\Lambda) +\frac{1}{2}\log|\Lambda| \\
    &+ \frac{n}{2}\log|\Lambda| + \frac{\nu_0 - d - 1}{2}\log|\Lambda|\bigg]\\
    &\propto -\frac{1}{2}\Tra\bigg( \big(\E{\Ypred}{\Ypred\Ypred^\top} + \E{\rho_\mu}{\mu\mu^\top} - 2\E{\rho_{\Ypred}}{\Ypred}\E{\rho_{\mu}}{\mu}^\top +\sum_{t=1}^n Y_t Y_t^\top -2\sum_{t=1}^n Y_t \E{\rho_{\mu}}{\mu}^\top \\
    &+ n\E{\rho_{\mu}}{\mu\mu^\top}+\psi_0^{-1} \big)\Lambda \bigg) + \frac{1}{2}(n+\nu_0-d-1)\log|\Lambda|\,.
\end{align*}
Identifying the corresponding terms with a Wishart distribution yields to
\begin{align*}
    \rho_\mu(\dint\Lambda) &= \cW(\dint\Lambda ; \nu_\Lambda, \psi_\Lambda)\\
    \text{where } &\nu_\Lambda = n+d+1\\
    &\psi_\Lambda^{-1} =\E{\Ypred}{\Ypred\Ypred^\top} + \E{\rho_\mu}{\mu\mu^\top} - 2\E{\rho_{\Ypred}}{\Ypred}\E{\rho_{\mu}}{\mu}^\top +\sum_{t=1}^n Y_t Y_t^\top -2\sum_{t=1}^n Y_t \E{\rho_{\mu}}{\mu}^\top + n\E{\rho_{\mu}}{\mu\mu^\top}+\psi_0^{-1} \,.
\end{align*}

\begin{lemma}[Objective function under stationary Gaussian-Wishart model]\label{lemma:objective_function_standard_GW}
Fir any $\delta\in\cD$, let $(\xi_y, \Lambda_y,\xi_\mu, \Lambda_\mu, \nu_\Lambda, \psi_\Lambda)$ be the parameters of the corresponding variational distribution $\rhoVB$ under the stationary Gaussian-Wishart model \eqref{eq:model_GW}. Then, the objective function can be written as
    \begin{align*}
        \obj(\delta) &= -\frac{\nu_\Lambda}{2}\mathrm{Tr}\left(\left( \sum_{t\in[n]}Y_t Y_t^\top - 2\left( \sum_{t\in[n]}Y_t + \xi_y \right)\xi_\mu^\top + (n+1)(\Lambda_\mu^{-1} + \xi_\mu\xi_\mu^\top) + \Lambda_y^{-1} + \xi_y\xi_y^\top + \psi_0^{-1}\right)\psi_\Lambda \right)\\
        &-\frac{1}{2}\mathrm{Tr}\big( ( \Lambda_\mu^{-1} + \xi_\mu\xi_\mu^\top)\Lambda_0 \big) + \xi_\mu^\top \Lambda_0\mu_0 + \frac{1}{2}(n+\nu_0+1)\log\det(\psi_\Lambda) - \frac{1}{2}\big(\log\det(\Lambda_y)+\log\det(\Lambda_\mu)\big) - \lambda\delta^\top\xi_y\,.
    \end{align*}
\end{lemma}

\begin{proof}
From \cref{lemma:rho_GW}, we found that the variational distribution for the GW model can be written as
\begin{align*}
\rhoVB(\dint(\Ypred, \mu, \Lambda)) = \cN(\Ypred ; :\xi_y, \Lambda_y^{-1})\cN(\mu ; \xi_\mu, \Lambda_\mu^{-1})\cW(\Lambda ; \nu_\Lambda, \psi_{\Lambda})\,.
\end{align*}
Starting from the definition of $\risk$, we have, for any $\delta\in\cD$,
\begin{align*}
    \risk(\delta) = -\E{\rho_y}{\log\rho_y(\Ypred)} -\E{\rho_\mu}{\log\rho_\mu(\mu)} - \E{\rho_\Lambda}{\log\rho_\Lambda(\Lambda)} - \lambda\delta^\top\E{\rho_y}{\Ypred} + \E{\rhoVB}{\log\pi\left(\Ypred, \mu, \Lambda\condi H_n\right)} +  C\,, 
\end{align*}
where $C$ is a constant that does not depend on $\delta$. We now aim at computing each of these terms. One can easily verify that
\begin{align*}
&-\E{\rho_y}{\log\rho_y(\Ypred)} \propto_\delta -\frac{1}{2}\log|\Lambda_y| &&-\E{\rho_\mu}{\log\rho_\mu(\mu)} \propto_\delta -\frac{1}{2}\log|\Lambda_\mu| &&&-\E{\rho_\Lambda}{\log\rho_\Lambda(\Lambda)} \propto_\delta \frac{d+1}{2}\log|\psi_\Lambda|\,.
\end{align*}
Moreover, 
\begin{align*}
    \E{\rhoVB}{\log\pi(\Ypred, \mu, \Lambda\condi H_n)}&\propto_\delta \E{\rhoVB}{\log\prod_{t=1}^n \cN(Y_t ; \mu, \Lambda^{-1})} + \E{\rhoVB}{\log \cN(\Ypred ; \mu, \Lambda^{-1})} + \E{\rhoVB}{\log \cN(\mu ; \mu_0, \Lambda_0^{-1})}\\
    &+ \E{\rhoVB}{\log \cW(\Lambda ; \nu_0, \psi_0}\,.
\end{align*}
We can compute each of these terms exactly the same way we did in the proof of \cref{lemma:rho_GW}, and combining these with the terms above give the desited expression.
\end{proof}

\subsection{ \texorpdfstring{$\algVB$}{Lg} for the AR Model}\label{appendix:AR}
\begin{lemma}[Solution of \eqref{eq:def_rho_vb} under AR(1) model]\label{lemma:rho_Ar}
Under \emph{AR} model \eqref{eq:model_AR1}, for any $\delta\in\cD$, the corresponding variational distribution $\rhoVB$ can be factorised as follows,
\begin{align*}
    \rhoVB(\dint(\Ypred, \Gamma, \Lambda)) =  \rho_y(\dint\Ypred)\rho_\Gamma(\dint(\ve(\Gamma)))\rho_\Lambda(\dint\Lambda)\,,
\end{align*}
where $\rho_y(\dint\Ypred) = \cN(\dint\Ypred;\xi_y, \Lambda_y^{-1})$, $\rho_\Gamma(\dint(\ve(\Gamma))) = \cN(\dint(\ve(\Gamma));\ve(M_\Gamma), V_\Gamma\otimes U_\Gamma)$, $\rho_\Lambda(\dint\Lambda) = \cW(\dint \Lambda ; \nu_\Lambda, \psi_\Lambda)$. Moreover, 
the variational parameters $\phi = (\xi_y, \Lambda_y,\ve(M_\Gamma), V_\Gamma \otimes U_\Gamma, \nu_\Lambda, \psi_\Lambda)$ satisfy a fixed-point equation $T_n(\phi) = \phi$, where $T_n$ is given as follows:
{\scriptsize\begin{align*}
T_n: \phi\mapsto \begin{pmatrix}
M_\Gamma Y_n - \frac{\lambda}{\nu_\Lambda} \psi_\Lambda^{-1}\delta\\
\nu_\Lambda\psi_\Lambda\\
(V_\Gamma \otimes U_\Gamma)\left[ \left( I_d \otimes \nu_\Lambda \psi_\Lambda \right)\ve(\sum_{t=1}^n Y_t Y_{t-1}^\top + \xi_y Y_n^\top) + (V_0^{-1}\otimes U_0^{-1})\ve(M_0) \right] \\
\left(\sum_{t=0}^n Y_t Y_t^\top \otimes \nu_\Lambda \psi_\Lambda + V_0^{-1}\otimes U_0^{-1}\right)^{-1}\\
n+\nu_0+1\\
\psi_0^{-1}+\xi_y \xi_y^\top + \Lambda_y^{-1} + \sum_{t=1}^n Y_t Y_t^\top - 2 M_\Gamma \left( \sum_{t\in[n]} Y_{t-1} Y_t^\top + Y_n \xi_y^\top \right) + M_\Gamma \sum_{t=0}^n Y_t Y_t^\top M_\Lambda^\top + \sum_{i=1}^{d^2} \sigma_i \ve^{-1}(u_i) \left(\sum_{t=0}^n Y_t Y_t^\top\right)\ve^{-1}(u_i)^\top
\end{pmatrix}\,,
\end{align*}}%
where ${(\sigma_i, u_i)}_{i\in[d]}$ is the spectral decomposition of $V_\Lambda \otimes U_\Lambda$.
\end{lemma}
\begin{proof}
First, we write $\Tilde{\pi}_n$ by remarking
\begin{align}\label{eq:AR1_pi_n_temp}
    \Tilde{\pi}_n(\Ypred, \Gamma, \Lambda) &\propto e^{-\lambda \delta^\top \Ypred}\pi(\Ypred, \Gamma, \Lambda \condi H_n)\nonumber\\
    &\propto e^{-\lambda \delta^\top \Ypred}\pi(H_n\condi Y_{n+1}, \Gamma, \Lambda)\pi(\Ypred, \Gamma, \Lambda)\nonumber\\
     &\propto e^{-\lambda \delta^\top \Ypred}\pi(H_n\condi Y_{n+1}, \Gamma, \Lambda)\pi(\Ypred\condi \Gamma, \Lambda)\pi_0(\Gamma, \Lambda)\,,
\end{align}
On behalf of \eqref{eq:model_AR1}, \eqref{eq:AR1_pi_n_temp} becomes
\begin{align*}
    \Tilde{\pi}_n(\Ypred, \Gamma, \Lambda) &\propto e^{-\lambda \delta^\top \Ypred} \left(\prod_{t=1}^n \cN(Y_t;\Gamma Y_{t-1}, \Lambda^{-1})\right)\pi(\Ypred;\Gamma Y_{t},  \Lambda^{-1})\mathcal{MN}(\Gamma;M_0, U_0, V_0)\cW(\Lambda ; \nu_0, \psi_0)\,.
\end{align*}
Keeping only terms that depend on the variable $\Ypred$,
\begin{align*}
    \log\rho_y(\Ypred) &\propto_{\Ypred}\E{\rho_\Gamma, \rho_\Lambda}{\log\left(e^{-\lambda\delta^\top\Ypred}\cN(\Ypred ; \Gamma Y_n, \Lambda^{-1})\right)}\\
    &\propto -\lambda \delta^\top \Ypred - \frac{1}{2}\left(\Ypred^\top \E{\rho_\Lambda}{\Lambda}\Ypred - 2\Ypred^\top \E{\rho_\Lambda}{\Lambda}\E{\rho_\Gamma}{\Gamma} Y_n \right)\\
    &\propto -\frac{1}{2}\left(\Ypred^\top\E{\rho_\Lambda}{\Lambda}\Ypred - 2\Ypred^\top \left(\E{\rho_\Lambda}{\Lambda}\E{\rho_\Gamma}{\Gamma}Y_n - \lambda\delta \right) \right)\,,
\end{align*}
which gives, by completying the Gaussian square with respect tp $\Ypred$,
\begin{align*}
    \rho_y(\dint\Ypred) &= \cN(\dint\Ypred ; \xi_y, \Lambda_y^{-1})\\
    \text{where }&\xi_y = \E{\rho_\Gamma}{\Gamma}\left(Y_n - \lambda \E{\rho_\Gamma}{\Gamma}^{-1}\E{\rho_\Lambda}{\Lambda}^{-1}\delta\right)\\
    &\Lambda_y = \E{\rho_\Lambda}{\Lambda}\,.
\end{align*}
For the variational distribution with respect to $\Gamma$,
\begin{align*}
\log\rho_\Gamma(\Gamma)&\propto_\Gamma \E{\rho_{\Ypred}, \rho_\Lambda}{ \log\left( \prod_{t=1}^n \cN(Y_t ; \Gamma Y_{t-1}, \Lambda^{-1})\cN(\Ypred;\Gamma Y_n, \Lambda^{-1}) \mathcal{MN}(\Gamma; M_0, U_0, V_0) \right)}\\
&\propto \underbrace{-\frac{1}{2}\sum_{t=1}^n (Y_t - \Gamma Y_{t-1})^\top \E{\rho_\Lambda}{\Lambda} (Y_t - \Gamma Y_{t-1})}_{\textcolor{blue}{(1)}} \underbrace{- \frac{1}{2}\E{\rho_{\Ypred}}{(\Ypred - \Gamma Y_n)^\top \E{\rho_\Lambda}{\Lambda} (\Ypred - \Gamma Y_n)}}_{\textcolor{blue}{(2)}} \\
& \underbrace{- \frac{1}{2}\left(\ve(\Gamma) - \ve(M_0)\right)^\top \left(V_0\otimes U_0\right)^{-1}\left(\ve(\Gamma) - \ve(M_0)\right)}_{\textcolor{blue}{(3)}}\,.
\end{align*}
\begin{align*}
    \textcolor{blue}{(1)}&\propto_\Gamma - \frac{1}{2}\sum_{t=1}^n \left( (\Gamma Y_{t-1})^\top \E{\rho_\Lambda}{\Lambda}\Gamma Y_{t-1} - 2Y_t^\top \E{\rho_\Lambda}{\Lambda}\Gamma Y_{t-1} \right)\\
    &\propto -\frac{1}{2}\left( \Tra\left(\Gamma^\top \E{\rho_\Lambda}{\Lambda} \Gamma \sum_{t=1}^n Y_{t-1}Y_{t-1}^\top\right) - 2\Tra\left(\sum_{t=1}^n Y_{t-1}Y_t^\top \E{\rho_\Lambda}{\Lambda}\Gamma\right) \right)\\
    &\propto -\frac{1}{2}\left( \ve(\Gamma)^\top \left(\sum_{t=1}^n Y_{t-1}Y_{t-1}^\top\otimes \E{\rho_\Lambda}{\Lambda}\right)\ve(\Gamma) - 2\ve\left( \sum_{t=1}^n Y_t Y_{t-1}^\top \right)^\top (I_d\otimes \E{\rho_\Lambda}{\Lambda})\ve(\Gamma) \right)\,,
\end{align*}
where the last line follows from \cref{prop:vectorization_properties}.
We can show with the exact same arguments that
\begin{align*}
    \textcolor{blue}{(2)}\propto_\Gamma -\frac{1}{2}\left( \ve(\Gamma)^\top \left(Y_n Y_n^\top\otimes \E{\rho_\Lambda}{\Lambda}\right)\ve(\Gamma) - 2\ve\left( \E{\rho_{\Ypred}}{\Ypred} Y_{n-1}^\top \right)^\top (I_d\otimes \E{\rho_\Lambda}{\Lambda})\ve(\Gamma) \right)\,,
\end{align*}
and for the third term,
\begin{align*}
    \textcolor{blue}{(3)}&\propto -\frac{1}{2}\left( \ve(\Gamma)^\top\left(V_0\otimes U_0\right)^{-1}\ve(\Gamma) - 2\ve(\Gamma)^\top \left(V_0\otimes U_0\right)^{-1}\ve(M_0) \right)\,.
\end{align*}
Adding \textcolor{blue}{(1)}, \textcolor{blue}{(2)} and \textcolor{blue}{(3)} and completing the corresponding Gaussian squares gives
\begin{align*}
    \log\rho_\Gamma(\dint\Gamma) &= \cN(\dint\ve(\Gamma) ; \xi_\Gamma, \Lambda_\Gamma^{-1})\\
    \text{where }&\Lambda_\Gamma = \sum_{t=0}^n Y_t Y_t^\top \otimes \E{\rho_\Lambda}{\Lambda} + V_0^{-1}\otimes U_0^{-1}\\
    &\xi_\Gamma = \Lambda_\Gamma^{-1}\left( \left(I_d\otimes \E{\rho_\Lambda}{\Lambda}\right)\left( \ve\left(\sum_{t=1}^n Y_t Y_{t-1}^\top\right) + \E{\rho_{\Ypred}}{\Ypred}Y_n^\top \right) + (V_0^{-1}\otimes U_0^{-1})\ve(M_0) \right)\,.
\end{align*}
The variational distribution with respect to $\Lambda$ requires more efforts:
\begin{align*}
    \log\rho_\Lambda(\Lambda)&\propto_\Lambda \E{\rho_{\Ypred}, \rho_{\Gamma}}{\log\prod_{t=1}^n \cN(Y_t;\Gamma Y_{t-1}, \Lambda^{-1})\cN(\Ypred ; \Gamma Y_n, \Lambda^{-1})\cW(\Lambda ; \nu_0, \psi_0)}\\
    &\propto \E{\rho_{\Ypred}, \rho_{\Gamma}}{ -\frac{1}{2}\sum_{t=1}^n (Y_t - \Gamma Y_{t-1})^\top \Lambda (Y_t - \Gamma Y_{t-1}) - \frac{1}{2}(\Ypred - \Gamma Y_n)^\top \Lambda (\Ypred - \Gamma Y_{n-1}) } + \frac{n + \nu_0 - d - 1}{2}\log|\Lambda|\\
    &- \frac{1}{2}\Tra(\Lambda\psi_0^{-1})\\
    &\propto -\frac{1}{2}\Tra\left(\Lambda\underbrace{\left( \E{\rho_\Gamma}{\sum_{t=1}^n (Y_t - \Gamma Y_{t-1})(Y_t - \Gamma Y_{t-1})^\top} + \E{\rho_{\Ypred}, \rho_\Gamma}{(\Ypred - \Gamma Y_{n})(\Ypred - \Gamma Y_{n})^\top} +\psi_0^{-1}\right)}_{\textcolor{blue}{(\star)}}\right)\\
    &+ \frac{n + \nu_0 - d - 1}{2}\log|\Lambda|\,.
\end{align*}
where we refer to the proof in \cref{lemma:rho_GW} for the computation in the second line. Then term inside the trace writes
\begin{align*}
\textcolor{blue}{(\star)} &= \sum_{t=1}^n Y_t Y_t^\top + \sum_{t=1}^n \E{\rho_\Gamma}{\Gamma Y_{t-1}Y_{t-1}\Gamma^\top} - \E{\rho_\Gamma}{\Gamma}\sum_{t=1}^n Y_{t-1}Y_t^\top - \sum_{t=1}^n Y_t Y_{t-1}^\top \E{\rho_\Gamma}{\Gamma}^\top+\psi_0^{-1} \\
&+ \E{\rho_{\Ypred}}{\Ypred\Ypred^\top} + \E{\rho_\Gamma}{\Gamma Y_n Y_n \Gamma^\top} - \E{\rho_\Gamma}{\Gamma}Y_n\E{\rho_{\Ypred}}{\Ypred}^\top - \E{\rho_{\Ypred}}{\Ypred}Y_n \Gamma^\top\\
&=\sum_{t=1}^n Y_t Y_t^\top + \sum_{t=1}^n \E{\rho_\Gamma}{\Gamma Y_{t-1}Y_{t-1}\Gamma^\top} - \E{\rho_\Gamma}{\Gamma}\sum_{t=1}^n Y_{t-1}Y_t^\top - \sum_{t=1}^n Y_t Y_{t-1}^\top \E{\rho_\Gamma}{\Gamma}^\top \\
&+ \E{\rho_{\Ypred}}{\Ypred}\E{\rho_{\Ypred}}{\Ypred}^\top + \cov_{\rho_{\Ypred}}(\Ypred) + \E{\rho_\Gamma}{\Gamma Y_n Y_n \Gamma^\top} - \E{\rho_\Gamma}{\Gamma}Y_n\E{\rho_{\Ypred}}{\Ypred}^\top - \E{\rho_{\Ypred}}{\Ypred}Y_n \Gamma^\top\,.
\end{align*}
The main difficulty here is to compute the term related to $\sum_{t=1}^n \E{\rho_\Gamma}{\Gamma Y_{t-1}Y_{t-1}\Gamma^\top} + \E{\rho_\Gamma}{\Gamma Y_n Y_n \Gamma^\top} = \sum_{t=0}^n \E{\rho_\Gamma}{\Gamma Y_t Y_t \Gamma^\top}$; extracting this term gives
\begin{align*}
    \Tra\left(\Lambda\sum_{t=0}^n \E{\rho_\Gamma}{\Gamma Y_t Y_t \Gamma^\top}\right) &= \E{\rho_\Gamma}{\Tra\left(\Gamma^\top \Lambda \Gamma\sum_{t=0}^n Y_t Y_t^\top\right)}\\
    &= \E{\rho_\Gamma}{ \ve(\Gamma)^\top \left(\sum_{t=0}^n Y_t Y_t^\top\otimes \Lambda\right)\ve(\Gamma) }\\
    &= \Tra\left( \left(\sum_{t=0}^n Y_t Y_t^\top\otimes \Lambda\right)\E{\rho_\Gamma}{\ve(\Gamma)\ve(\Gamma)^\top} \right)\\
    &=\underbrace{\Tra\left( \left(\sum_{t=0}^n Y_t Y_t^\top \otimes \Lambda\right)\ve(\E{\rho_{\Gamma}}{\ve(\Gamma)} )\ve(\E{\rho_{\Gamma}}{\ve(\Gamma)} )^\top\right)}_{\textcolor{blue}{(1)}} \\
    &+\underbrace{\Tra\left( \left(\sum_{t=0}^n Y_t Y_t^\top \otimes \Lambda\right) \cov_{\rho_\Gamma}\left(\ve(\Gamma)\right) \right)}_{\textcolor{blue}{(2)}}\,.
\end{align*}
The first trace term of the last line writes
\begin{align*}
    \textcolor{blue}{(1)} &= \E{\rho_{\Gamma}}{\ve(\Gamma)}^\top\left(\sum_{t=0}^n Y_t Y_t^\top \otimes \Lambda\right)\E{\rho_{\Gamma}}{\ve(\Gamma)} = \Tra\left(  \E{\rho_{\Gamma}}{\Gamma}^\top \Lambda  \E{\rho_{\Gamma}}{\Gamma} \sum_{t=0}^n Y_t Y_t^\top \right)= \Tra\left(  \Lambda \E{\rho_{\Gamma}}{\Gamma} \sum_{t=0}^n Y_t Y_t^\top   \E{\rho_{\Gamma}}{\Gamma}^\top\right)\,,
\end{align*}
while the second term gives, 
\begin{align*}
     \textcolor{blue}{(2)} &= \Tra\left( \left(\sum_{t=0}^n Y_t Y_t^\top \otimes \Lambda\right)\sum_{i=1}^{d^2} \sigma_i u_i u_i^\top \right)\,,
\end{align*}
where we denote ${(\sigma_i, u_i)}_{i=1}^{d^2}$ the spectral decomposition of the covariance matrix $\cov_{\rho_\Gamma}\left(\ve(\Gamma)\right)$. Then we have
\begin{align*}
     \textcolor{blue}{(2)} = \sum_{i=1}^{d^2} \sigma_i u_i^\top \left(\sum_{t=0}^n Y_t Y_t^\top \otimes \Lambda\right)u_i &= \sum_{i=1}^{d^2}\sigma_i \Tra\left(\ve^{-1}(u_i)^\top \Lambda \ve^{-1}(u_i) \sum_{t=0}^n Y_t Y_t^\top \right) \\
     &= \sum_{i=1}^{d^2}\sigma_i \Tra\left(\Lambda \ve^{-1}(u_i) \sum_{t=0}^n Y_t Y_t^\top\ve^{-1}(u_i)^\top \right)\,.
\end{align*}
Combining all these yields to
\begin{align*}
    \rho_{\Lambda}(\dint\Lambda) &= \cW(\dint\Lambda ; \nu_\Lambda, \psi_\Lambda)\\
    \text{where }&\nu_\Lambda = n + d + 1\\
    &\psi_\Lambda = \psi_0^{-1} + \sum_{t=1}^n Y_t Y_t^\top + \E{\rho_{\Ypred}}{\Ypred}\E{\rho_{\Ypred}}{\Ypred}^\top + \cov_{\rho_{\Ypred}}(\Ypred) - \E{\rho_\Gamma}{\Gamma}Y_n\E{\rho_{\Ypred}}{\Ypred}^\top \\
    &- \E{\rho_{\Ypred}}{\Ypred}Y_n \E{\rho_\Gamma}{\Gamma}^\top- \E{\rho_\Gamma}{\Gamma}\sum_{t=1}^n Y_{t-1}Y_t^\top- \sum_{t=1}^n Y_t Y_{t-1}^\top \E{\rho_\Gamma}{\Gamma}^\top\\
    &+\E{\rho_{\Gamma}}{\Gamma} \sum_{t=0}^n Y_t Y_t^\top   \E{\rho_{\Gamma}}{\Gamma}^\top + \sum_{i=1}^{d^2}\sigma_i\ve^{-1}(u_i) \sum_{t=0}^n Y_t Y_t^\top\ve^{-1}(u_i)^\top \,.
\end{align*}
\end{proof}
We next derive the corresponding objective function.
\begin{lemma}[Objective function under AR model]\label{lemma:objective_function_AR1}
For any $\delta\in\cD$, let $(\xi_y, \Lambda_y,M_\Gamma, V_\Gamma\otimes U_\Gamma, \nu_\Lambda, \psi_\Lambda)$ the parameters of the corresponding variational distribution $\rhoVB$ under the \emph{AR} model \eqref{eq:model_AR1}. Then, the objective function can is written as
\begin{align*}
    \obj(\delta) &= -\frac{\nu_\Lambda}{2}\mathrm{Tr}\bigg(\Big(\psi_0^{-1}+\xi_y\xi_y^\top + \Lambda_y^{-1} + \sum_{t=1}^n Y_t Y_t^\top - 2 M_\Gamma \left(\sum_{t=1}^n Y_{t-1} Y_{t}^\top + Y_n \xi_y^\top\right)M_\Lambda^\top +M_\Lambda \sum_{t=0}^n Y_t Y_t^\top M_\Lambda^\top + \\
    &\sum_{i=1}^{d^2}\sigma_i \ve^{-1}(u_i)\sum_{t=0}^n Y_t Y_t^\top \ve^{-1}(u_i)^\top \Big)\psi_\Lambda\bigg)-\frac{1}{2}\mathrm{Tr}\left( (V_0^{-1}\otimes U_0^{-1})\left(\ve(M_\Lambda)\ve(M_\Lambda)^\top + V_\Lambda\otimes U_\Lambda\right) \right) \\
    &+ \ve(M_\Lambda)^\top \left(V_0\otimes U_0\right)^{-1}\ve(M_0)+\frac{1}{2}(n+\nu_0+1)\log\det(\psi_\Lambda)-\frac{1}{2}\log\det(\Lambda_y) - \frac{1}{2}\log\det(V_\Lambda\otimes U_\Lambda)-\lambda\delta^\top \xi_y\,.
\end{align*}
\end{lemma}

\begin{proof}
From \cref{lemma:rho_GP}, we found that the variational distribution for the AR model can be written as

\begin{align*}
    \rhoVB(\dint(\Ypred, \Gamma, \Lambda)) =  \cN(\dint\Ypred;\xi_y, \Lambda_y^{-1})\cN(\dint(\ve(\Gamma));\ve(M_\Gamma), V_\Gamma\otimes U_\Gamma) \cW(\dint \Lambda ; \nu_\Lambda, \psi_\Lambda)\,.
\end{align*}
Starting again from the definition of $\risk$, we have, for any $\delta\in\cD$,
\begin{align*}
    \risk(\delta) \propto_\delta -\E{\rho_y}{\log\rho_y(\Ypred)} -\E{\rho_\Gamma}{\log\rho_\Gamma(\Gamma)} - \E{\rho_\Lambda}{\log\rho_\Lambda(\Lambda)} - \lambda\delta^\top\E{\rho_y}{\Ypred} + \E{\rhoVB}{\log\pi\left(\Ypred, \Gamma, \Lambda\condi H_n\right)}\,, 
\end{align*}
where
\begin{align*}
-\E{\rho_y}{\log\rho_y(\Ypred)} \propto_\delta -\frac{1}{2}\log|\Lambda_y| &&-\E{\rho_\Gamma}{\log\rho_\Gamma(\Gamma)} \propto_\delta -\frac{1}{2}\log|V_\Lambda \otimes U_\Lambda| &&&-\E{\rho_\Lambda}{\log\rho_\Lambda(\Lambda)} \propto_\delta \frac{d+1}{2}\log|\psi_\Lambda|\,.
\end{align*}
Moreover, 
\begin{align*}
    \E{\rhoVB}{\log\pi(\Ypred, \Gamma, \Lambda\condi H_n)}&\propto_\delta \E{\rhoVB}{\log\prod_{t=1}^n \cN(Y_t ; \Gamma Y_{t-1}, \Lambda^{-1})} + \E{\rhoVB}{\log \cN(\Ypred ; \Gamma Y_n, \Lambda^{-1})} \\
    &+ \E{\rhoVB}{\log \mathcal{MN}(\Gamma ; M_0,U_0, V_0)} + \E{\rhoVB}{\log \cW(\Lambda ; \nu_0, \psi_0}\,.
\end{align*}
We can compute each of these terms exactly the same way we did in the proof of \cref{lemma:rho_Ar}, and combining these with the terms above give the desited expression.    
\end{proof}

\subsection{ \texorpdfstring{$\algVB$}{Lg} for the Gaussian Process Model}\label{Appendix:GP}
\begin{lemma}[Solution of \eqref{eq:def_rho_vb} under Gaussian-process Wishart model]\label{lemma:rho_GP}
Under \emph{GP} model \eqref{eq:model_GW}, for any $\delta\in\cD$, the corresponding variational distribution $\rhoVB$ can be factorised as follows,
\begin{align*}
    \rhoVB(\dint(\Ypred, \mu, \Lambda)) =  \rho_y(\dint\Ypred)\rho_\mu(\dint\mu)\rho_\Lambda(\dint\Lambda)\,,
\end{align*}
where $\rho_y(\dint\Ypred) = \cN(\dint\Ypred;\xi_y, \Lambda_y^{-1})$, $\rho_\mu(\dint\mu) = \MGP(\dint\mu;m_\mu(\cdot), k_\mu(\cdot), \Omega_\mu)$, $\rho_\Lambda(\dint\Lambda) = \cW(\dint \Lambda ; \nu_\Lambda, \psi_\Lambda)$. At time step $n+1$, the variational parameters $\phi_{n+1} = (\xi_y, \Lambda_y,m_\mu^{1:n+1}, (\Omega_\mu \otimes K_\mu)^{1:n+1}, \nu_\Lambda, \psi_\Lambda)$ satisfy a fixed-point equation $T_n(\phi_{n+1}) = \phi_{n+1}$, where $T_n$ is given as follows:
{\tiny\begin{align*}
T_n: \phi\mapsto \begin{pmatrix}
m_\mu(n+1)-\frac{\lambda}{\nu_\Lambda}\psi_\Lambda^{-1}\delta\\
\nu_\Lambda \psi_\Lambda\\
(\Omega_\mu \otimes K_\mu)\left( \left(I_{n+1}\otimes \nu_\Lambda \psi_\Lambda \right)\ve(\bm{Y}) + \left(\Omega_0^{-1}\otimes K_0^{-1}\right)\ve(M_0) \right)\\
\left( I_{n+1}\otimes\nu_\Lambda\psi_\Lambda + \Omega_0^{-1}\otimes K_0^{-1} \right)^{-1}\\
\nu_0+n+1\\
\psi_0^{-1}+\xi_y\xi_y^\top + \Lambda_y^{-1} + \sum_{t\in[n]}Y_t Y_t^\top - 2\sum_{t\in[n]}m_\mu(t)Y_t^\top - 2m_\mu(n+1)\xi_y^\top +\sum_{t=1}^{n+1}m_\mu(t)m_\mu(t)^\top + \sum_{t=1}^{n+1}\Cov(\mu(t), \mu(t))
\end{pmatrix}\,.
\end{align*}}%
\end{lemma}

\begin{proof}
First, we write $\Tilde{\pi}_n$ as
\begin{align*}
    \Tilde{\pi}_n(\Ypred, \mu, \Lambda)&\propto e^{-\lambda \delta^\top \Ypred}\pi(\Ypred, \mu, \Lambda \condi H_n) \\
    &\propto e^{-\lambda \delta^\top \Ypred}\pi(H_n\condi \mu, \Lambda)\pi(\Ypred\condi \mu, \Lambda)\pi_0(\mu, \Lambda)\\
    &\propto e^{-\lambda \delta^\top \Ypred}\prod_{t=1}^n \cN(Y_t ; \mu(t), \Lambda^{-1})\cN(\Ypred ;\mu(t+1), \Lambda^{-1})\MGP(\mu ; \mu_0(\cdot), K_0, \Omega_0)\cW(\Lambda ; \nu_0, \psi_0)\,.
\end{align*}

First, the variational distribution $\rho_y$ can be derived as
\begin{align*}
    \log\rho_y(\Ypred) &\propto_{\Ypred} \E{\rho_\mu, \rho_\Lambda}{ \log e^{-\lambda \delta^\top \Ypred}\cN(\Ypred ; \mu(t+1), \Lambda^{-1}) }\\
    &\propto \E{\rho_\mu, \rho_\Lambda}{ -\lambda \delta^\top \Ypred - \frac{1}{2}\left(\Ypred - \mu(t+1)\right)^\top \Lambda \left(\Ypred - \mu(t+1)\right) }\\
    &\propto -\frac{1}{2}\left( \Ypred^\top \Lambda \Ypred - 2\Ypred^\top \left(\E{\rho_\Lambda}{\Lambda}\E{\rho_\mu}{\mu(t+1)} - \lambda \delta\right) \right)\\
    &\propto \log\cN\left(\Ypred ; \xi_y, \Lambda_y^{-1}\right)\,.
\end{align*}

The variational distribution $\rho_y$ can be derived by deploying the GP prior on the indices $\{1,\dots n+1\}$;
\begin{align*}
    &\log\rho_\mu(\mu)\\
    &\propto_\mu \E{\rho_y, \rho_\Lambda}{ \log\prod_{t=1}^n \cN(Y_t ; \mu, \Lambda^{-1})\cN(\Ypred ; \mu, \Lambda^{-1})\MGP(\mu ; \mu_0(\cdot), K_0, \Omega_0) }\\
    &\propto \mathbb{E}_{\rho_y, \rho_\Lambda}\bigg[-\frac{1}{2}\sum_{t=1}^n \left(Y_t - \mu(t)\right)^\top \Lambda \left(Y_t - \mu(t)\right) \\
    &-\frac{1}{2}\left(\Ypred - \mu(n+1)\right)^\top \Lambda \left(\Ypred - \mu(n+1)\right) - \frac{1}{2}\mujoint^\top \left(\Omega_0\otimes \Kjoint \right)^{-1}\mujoint  \bigg]\,,
\end{align*}
where we define $\mujoint$ as the concatenated vector $(\mu(1),\dots,\mu(n+1))$ of size $(n+1)\times d$, and $\Kjoint$ as the matrix of size $(n+1, n+1)$ whose entries are $K_0(i, j)$ for $i, j\in[n+1]$. Then we have
\begin{align*}
\log\rho_\mu(\mu)&\propto_\mu -\frac{1}{2}\bigg( \sum_{t=1}^n \mu(t)^\top \E{\rho_\Lambda}{\Lambda}\mu(t) - 2\sum_{t=1}^n \mu(t)^\top \E{\rho_\Lambda}{\Lambda} Y_t +\mu(n+1)^\top \E{\rho_\Lambda}{\Lambda} \mu(n+1) - 2\mu(n+1)^\top \E{\rho_\Lambda}{\Lambda} \E{\rho_y}{\Ypred} \\
& +\mujoint\left(\Omega_0\otimes \Kjoint\right)^{-1} - 2\mujoint \left(\Omega_0\otimes \Kjoint\right)^{-1}\mujoint \bigg)\\
&\propto -\frac{1}{2}\bigg(\mujoint^\top \left(I_{n+1}\otimes \E{\rho_\Lambda}{\Lambda} + \left(\Omega_0\otimes \Kjoint\right)^{-1} \right)\mujoint \\
&- 2\mujoint\left( \left(\Omega_0\otimes \Kjoint\right)^{-1}\mupriorjoint + \left(I_{n+1}\otimes\E{\rho_\Lambda}{\Lambda} \right)\bm{Y}_{1:n+1} \right)\bigg)\,,
\end{align*}
where we define $\mupriorjoint$ as the concatenated vector $(\mu_0(1),\dots,\mu_0(n+1))$, and $\bm{Y}_{1:n+1}$ the concatenated vector $\left(Y_1,\dots,Y_n, \xi_y\right)$. Therefore, we have
\begin{align*}
    \log\rho_\mu(\mu)\propto \log\MGP(\mu ; m_\mu(\cdot), \Omega_\mu, K_\mu)\,,
\end{align*}
where we define the mean function and covariance functions on indices $\{1,\dots n+1\}$,
\begin{align*}
    &m_\mu^{1:n+1} = \left( \Omega_\mu\otimes K_\mu^{1:n+1} \right) \left( \left(\Omega_0\otimes \Kjoint\right)^{-1}\mupriorjoint + \left(I_{n+1}\otimes\E{\rho_\Lambda}{\Lambda} \right)\bm{Y}_{1:n+1} \right)\\
    & \left( \Omega_\mu\otimes K_\mu^{1:n+1} \right)^{-1} = \left(I_{n+1}\otimes \E{\rho_\Lambda}{\Lambda} + \left(\Omega_0\otimes \Kjoint\right)^{-1} \right)\,. 
\end{align*}
Finally, the variational distribution $\rho_\Lambda$ can be derived as
\begin{align*}
    \log\rho_\Lambda(\Lambda)&\propto_\Lambda \E{\rho_y, \rho_\mu}{ \prod_{t=1}^n \cN(Y_y ; \mu(t), \Lambda^{-1}) \cN(\Ypred ; \mu(n+1), \Lambda^{-1})\cW(\Lambda ; \nu_0, \psi_0)  }\\
    &\propto \underbrace{\E{\rho_y, \rho_\mu}{ -\frac{1}{2}\sum_{t=1}^n\left(Y_t - \mu(t)\right)^\top \Lambda \left(Y_t - \mu(t)\right) - \frac{1}{2}\left(\Ypred - \mu(n+1)\right)^\top \Lambda \left(\Ypred - \mu(n+1)\right)}}_{\textcolor{blue}{(*)}} - \frac{1}{2}\Tra(\Lambda\psi_0^{-1}) \\
    &+ \frac{\nu_0 + n - d - 1}{2}\log|\Lambda|\,.
\end{align*}
The term inside the expectation gives
\begin{align*}
    \textcolor{blue}{(*)} &\propto_{\Lambda}-\frac{1}{2}\Tra\bigg( \Lambda\bigg(\sum_{t=1}^n Y_t Y_t^\top - 2\sum_{t=1}^n \E{\rho_\mu}{\mu(t)}Y_t^\top + \sum_{t=1}^n \E{\rho_\mu}{\mu(t)\mu(t)^\top} + \E{\rho_y}{\Ypred\Ypred^\top} - 2\E{\rho_y}{\Ypred}\E{\rho_\mu}{\mu(n+1)} \\
    &+ \E{\rho_\mu}{\mu(n+1)\mu(n+1)^\top} \bigg) \bigg)\\
    &\propto -\frac{1}{2}\Tra\bigg( \Lambda\bigg(\sum_{t=1}^n Y_t Y_t^\top - 2\sum_{t=1}^n m_\mu(t)Y_t^\top + \sum_{t=1}^{n+1} m_\mu(t)m_\mu(t) + \xi_y \xi_y^\top + \Lambda_y^{-1}- 2\xi_ym_\mu(n+1) + \sum_{t=1}^{n+1} \Cov(\mu(t), \mu(t)) \bigg) \bigg)\,.
\end{align*}
Hence, 
\begin{align*}
    \log\rho_\Lambda(\Lambda)\propto\log\cW(\Lambda ; \nu_\Lambda, \psi_\Lambda)\,,
\end{align*}
where
\begin{align*}
    &\nu_\Lambda = \nu_0 + n + 1\\
    &\psi_\Lambda = \psi_0^{-1}+\xi_y\xi_y^\top + \Lambda_y^{-1} + \sum_{t\in[n]}Y_t Y_t^\top - 2\sum_{t\in[n]}m_\mu(t)Y_t^\top - 2m_\mu(n+1)\xi_y^\top +\sum_{t=1}^{n+1}m_\mu(t)m_\mu(t)^\top + \sum_{t=1}^{n+1}\Cov(\mu(t), \mu(t))\,.
\end{align*}
\end{proof}

\begin{lemma}[Objective function under Gaussian Process model]\label{lemma:objective_function_GP}
For any $\delta\in\cD$, let $(\xi_y, \Lambda_y,m_\mu^{1:n+1}, (\Omega_\mu \otimes K_\mu)^{1:n+1}, \nu_\Lambda, \psi_\Lambda)$ the parameters of the corresponding variational distribution $\rhoVB$ under the \emph{GP} model \eqref{eq:model_GP}. Then, the objective function can is written as
\begin{align*}
&\obj(\delta) = -\frac{\nu_\Lambda}{2}\mathrm{Tr}\bigg( \bigg( \psi_0^{-1}+\xi_y\xi_y^\top + \Lambda_y^{-1} + \sum_{t\in[n]}Y_t Y_t^\top - 2\sum_{t\in[n]}m_\mu(t)Y_t^\top - 2m_\mu(n+1)\xi_y^\top \\
&+\sum_{t=1}^{n+1}m_\mu(t)m_\mu(t)^\top + \sum_{t=1}^{n+1}\Cov(\mu(t), \mu(t)) \bigg)\psi_\Lambda \bigg) - \frac{1}{2}\mathrm{Tr}\left( \Omega\otimes K_0^{n+1} \right)^{-1}\left( m_\mu^{1:n+1} (m_\mu^{1:n+1})^\top + \Omega_\mu \otimes K_\mu^{n+1}\right)\\
&+ (m_\mu^{1:n+1})^\top\left(\Omega_0\otimes K_0^{n+1}\right)^{-1}\mu_0^{1:n+1} + \frac{1}{2}(n+\nu_0+1)\log\det(\psi_\Lambda) - \frac{1}{2}\left( \log\det(\Lambda_y) + \log\det(\Omega_\mu\otimes K_\mu^{1:n+1}) \right)- \lambda\delta^\top\xi_y 
\end{align*}
\end{lemma}

\begin{proof}
From \cref{lemma:rho_GP}, we found that the variational distribution for the AR model can be written as

\begin{align*}
    \rhoVB(\dint(\Ypred, \mu, \Lambda)) =  \cN(\dint\Ypred;\xi_y, \Lambda_y^{-1})\MGP(\dint\mu;m_\mu(\cdot), k_\mu(\cdot), \Omega_\mu)\cW(\dint \Lambda ; \nu_\Lambda, \psi_\Lambda)\,,
\end{align*}

Starting again from the definition of $\risk$, we have, for any $\delta\in\cD$,
\begin{align*}
    \risk(\delta) \propto_\delta -\E{\rho_y}{\log\rho_y(\Ypred)} -\E{\rho_\mu}{\log\rho_\Gamma(\mu)} - \E{\rho_\Lambda}{\log\rho_\Lambda(\Lambda)} - \lambda\delta^\top\E{\rho_y}{\Ypred} + \E{\rhoVB}{\log\pi\left(\Ypred, \mu, \Lambda\condi H_n\right)} \,, 
\end{align*}
where
\begin{align*}
&-\E{\rho_y}{\log\rho_y(\Ypred)} \propto_\delta -\frac{1}{2}\log|\Lambda_y| &&-\E{\rho_\mu}{\log\rho_\mu(\mu)} \propto_\delta -\frac{1}{2}\log|\Omega_\mu\otimes K_\mu^{1:n+1}| &&&-\E{\rho_\Lambda}{\log\rho_\Lambda(\Lambda)} \propto_\delta \frac{d+1}{2}\log|\psi_\Lambda|\,.
\end{align*}
Moreover, 
\begin{align*}
    \E{\rhoVB}{\log\pi(\Ypred, \mu, \Lambda\condi H_n)}&\propto_\delta \E{\rhoVB}{\log\prod_{t=1}^n \cN(Y_t ; \Gamma \mu(t), \Lambda^{-1})} + \E{\rhoVB}{\log \cN(\Ypred ; \mu(n+1), \Lambda^{-1})} \\
    &+ \E{\rhoVB}{\log \mathcal{MGP}(\mu ; \mu_0,K_0, \Omega_0)} + \E{\rhoVB}{\log \cW(\Lambda ; \nu_0, \psi_0}\,.
\end{align*}
We can compute each of these terms exactly the same way we did in the proof of \cref{lemma:rho_GP}, and combining these with the terms above give the desited expression.    
\end{proof}

\subsection{Additional Models}
\textbf{AR(p) Model.} We can extend our AR model (which is in reality an AR(1) model) to an AR(p) model with $p\geq 2$ by extending the simply dimension. In fact, an AR(p) model writes 
\begin{align*}
    &Y_t \condi Y_{t-1},\dots  Y_{t-p}, (\Gamma_i)_{1\leq i\leq p}, \Lambda \sim \cN\left(\sum_{i=1}^p \Gamma_i Y_{t-i}, \Lambda\right)&&\forall t > p\nonumber\\
    &\Gamma_i \sim \cMN(M_0^i, U_0^i, V_0^i)&&\forall i\in[p]\\
    &\Lambda\sim\cW(\nu_0, \psi_0)\,,
\end{align*}
which can be written as a tensor AR(1) model,
\begin{align*}
    &\begin{pmatrix}
        Y_t\\
        Y_{t-1}\\
        \vdots\\
        Y_{t-p+1}
    \end{pmatrix}\condi (\Gamma_i)_{1\leq i\leq p}, \Lambda \sim \cN\left(C_{\left(\Gamma_1,\dots\Gamma_p\right)}\begin{pmatrix}
        Y_{t-1}\\
        Y_{t-2}\\
        \vdots\\
        Y_{t-p}
    \end{pmatrix}, \begin{pmatrix}
        \Lambda & 0 & \hdots & 0\\
        0 & 0 &\hdots & 0\\
        \vdots & \vdots &\ddots & \vdots\\
         0 & 0 &\hdots & 0\\
    \end{pmatrix}\right)\\
    &\Gamma_i \sim \cMN(M_0^i, U_0^i, V_0^i)\qquad\forall i\in[p]\\
    &\Lambda\sim\cW(\nu_0, \psi_0)\,,
\end{align*}
where
\begin{align*}
    C_{\left(\Gamma_1,\dots\Gamma_p\right)}=\begin{pmatrix}
        \Gamma_1 & \Gamma_2 & \hdots & \Gamma_{p-1} & \Gamma_{p}\\
        I_d & 0 &\hdots & 0& 0\\
        0 & I_d &\hdots & 0& 0\\
        \vdots & \vdots &\ddots & \vdots& \vdots\\
         0 & 0 &\hdots & I_d& 0\\
    \end{pmatrix}\in\real^{dp, dp}
\end{align*}
is called the \emph{companion} matrix. Hence, the AR(p) model reduces to an AR(1) model on $Z_t = \left(Y_t,Y_{t-1},\dots Y_{t-p+1}\right)^\top$ with $p$ different matrix parameters to infer.

\section{Derivation of \MCMC for Specific Models}\label{app:MCMC_specific}
We derive specific instances of \cref{alg:MCMC_GW} for both the GW and AR models, with a particular focus on detailing the form of the conditional posteriors.
\subsection{GW Model}
The joint parameter posterior $\pi(\mu, \Lambda \condi H_n)$ cannot be derived in closed-form, but 
we have the following conditional posteriors
\begin{align*}
    &\pi(\dint\mu\condi \Lambda, H_n) = \cN\left(\dint\mu \;;\; \left( \Lambda + \frac{1}{n}\Lambda_0 \right)^{-1}\left(\Lambda \frac{1}{n}\sum_{t=1}^n Y_t + \frac{1}{n}\Lambda_0 \mu_0\right), \frac{1}{n}\left( \Lambda + \frac{1}{n}\Lambda_0 \right)^{-1} \right) \\
    &\pi(\dint\Lambda \condi \mu, H_n) = \cW\left(\dint\Lambda\condi \mu, H_n\right) = \cW\left(\dint\Lambda\;;\;n + \nu_0, \left( \sum_{t=1}^n\left(Y_t - \mu\right)\left(Y_t - \mu\right)^\top + \psi_0^{-1} \right)^{-1} \right) \,.
\end{align*}
Applying Gibbs sampling with these two conditional posteriors yield to a chain $(\mu^{(k)}, \Lambda^{(k)})_{k=1}^M$. For a given $\delta$, the distribution $\check{\pi}_k$ is defined as
\begin{align*}
    \check{\pi}_k(\dint \Ypred ) = \cN\left(\dint\Ypred ; \mu^{(k)} - \lambda\Sigma^{(k)}\delta , \Sigma^{(k)}\right)\,.
\end{align*}
Hence, the algorithm $\MCMC$(GW) is defined as follows:
\begin{algorithm}
\caption{\MCMC(GW): Portfolio Construction with MCMC for GW model.}
\label{alg:MCMC_GW_model}
\textbf{Input:} Dataset $H_n$, initial decision $\hat\delta^{(0)}$, number of Monte-Carlo samples $M$, risk parameter $\lambda$, step-size $\eta$, initial parameters $(\mu^{(0)}, \Lambda^{(0)})$.\\
\While{Not converging}{
\For{$k=1,\dots,M$}{$\mu^{(k)}\sim \pi(\dint\mu\condi \Lambda^{(k-1)}, H_n)$ and $\Lambda^{(k)}\sim \pi(\dint\Lambda\condi \mu^{(k)}, H_n)$.}
For all $k \in [M]$, sample $z^{(k)} \sim \cN\left(\dint\Ypred ; \mu^{(k)} - \lambda\Sigma^{(k)}\delta , \Sigma^{(k)}\right)$.\\
$\hat\delta^{(k+1)} \gets \mathrm{Proj}_{\cD}\left(\hat\delta^{(k)} + \eta\lambda\frac{1}{M} \sum_{k \in [M]} z^{(k)}\right)$}
Return $\hat\delta^{(\infty)} = \hat\delta^{\rm{MCMC}}$.
\end{algorithm}

\subsection{AR Model}
Here again, the joint posterior distribution $\pi(\Gamma, \Lambda\condi H_n)$ cannot be computed in closed-form; we can compute the conditional posterior as
\begin{align*}
    \pi(\dint\Gamma\mid H_n, \Lambda)&\propto \pi(H_n\mid \Gamma,\Lambda)\pi(\dint\Gamma)\\
    &\propto\prod_{t=1}^n \exp\left( -\frac{1}{2}\left((\Gamma Y_{t-1})^\top \Lambda (\Gamma Y_{t-1}) - 2(\Gamma Y_{t-1})^\top \Lambda Y_t\right) \right)\pi(\dint\Gamma)\\
    &\propto \exp\left( -\frac{1}{2}\mathrm{Tr}\left(\Gamma^\top\Lambda\Gamma G_n - 2\Gamma^\top \Lambda\sum_{t=1}^n Y_t Y_{t-1}\right) \right)\pi(\dint\Gamma)\\
    &\propto \exp\Bigg( -\frac{1}{2}\left(\ve(\Gamma)^\top (G_n \otimes \Lambda)\ve(\Gamma) - 2\ve(\Gamma)^\top (I\otimes \Lambda)\ve(\sum_{t=1}^n Y_t Y_{t-1})\right)\\
    &\times\exp\left(-\frac{1}{2}\left(\ve(\Gamma)^\top (G_n \otimes \Lambda)^{-1}\ve(\Gamma)\right)-2\ve(\Gamma)^\top (V_0\otimes U_0)^{-1}\ve(M_0)\right) \Bigg)\\
    &\propto \exp\bigg( -\frac{1}{2}\bigg(\ve(\Gamma)^\top \left((G_n \otimes \Lambda)^{-1}+(V_0\otimes U_0)^{-1}\right)\\
    &-2\Ve(\Gamma)^\top\left( (I\otimes \Lambda)\ve(\sum_{t=1}^n Y_t Y_{t-1}) + (V_0\otimes U_0)^{-1}\ve(M_0)\right) \bigg) \bigg)
\end{align*}

Therefore, we have $\pi(\dint\Gamma\condi H_n, \Lambda) = \cN\left(\dint\ve(\Gamma)\;;\; \mu_{\Gamma}(\Lambda), \Sigma_\Gamma(\Lambda)\right)$, where
\begin{align*}
    &\Sigma_{\Gamma}(\Lambda) =  \left((G_n \otimes \Lambda)^{-1}+(V_0\otimes U_0)^{-1}\right)^{-1}\\
    &\mu_{\Gamma}(\Lambda) = \Sigma_{\Gamma}(\Lambda)\left( (I\otimes \Lambda)\ve(\sum_{t=1}^n Y_t Y_{t-1}) + (V_0\otimes U_0)^{-1}\ve(M_0)\right)\,.
\end{align*}
We also have the conditional posterior for the precision matrix, 
\begin{align*}
    \pi\left(\dint\Lambda \condi \Gamma, H_n) = \mathcal{W}(\dint \Lambda\;;\; n + \nu_0, \left(\sum_{t=1}^n (Y_t - \Gamma Y_{t-1})(Y_t - \Gamma Y_{t-1})^\top + \psi_0^{-1}\right)^{-1} \right)
\end{align*}
Therefore, we can have samples from the joint posterior distribution $\pi(\d(\Gamma, \Lambda)\mid H_n)$. For a given sample $k$, conditionally on the posterior samples $(\Gamma^{(k)}, \Lambda^{(k)})_{k=1}^M$, and the distribution $\check{\pi}_k$ is given by
\begin{align*}
\Breve{\pi}_k(\dint\Ypred) \propto e^{-\lambda^\top \Ypred}\cN(\dint\Ypred\;;\;\Gamma^{(k)}Y_n, (\Lambda^{(k)})^{-1}) \propto \cN\left(\Ypred\;;\;\Gamma^{(k)}Y_n - \lambda (\Lambda^{(k)})^{-1}\delta, (\Lambda^{(k)})^{-1} \right)
\end{align*}
The algorithm $\MCMC$(AR) can be instantiated as follows:

\begin{algorithm}
\caption{\MCMC(AR): Portfolio Construction with MCMC for AR model.}
\label{alg:MCMC_AR_model}
\textbf{Input:} Dataset $H_n$, initial decision $\hat\delta^{(0)}$, number of Monte-Carlo samples $M$, risk parameter $\lambda$, step-size $\eta$, initial parameters $(\Gamma^{(0)}, \Lambda^{(0)})$.\\
\While{Not converging}{
\For{$k=1,\dots,M$}{$\Gamma^{(k)}\sim \pi(\dint\Gamma\condi \Lambda^{(k-1)}, H_n)$ and $\Lambda^{(k)}\sim \pi(\dint\Lambda\condi \Gamma^{(k)}, H_n)$.}
For all $k \in [M]$, sample $z^{(k)} \sim \cN\left(\dint\Ypred ; \Gamma^{(k)}Y_n - \lambda\Sigma^{(k)}\delta , \Sigma^{(k)}\right)$.\\
$\hat\delta^{(k+1)} \gets \mathrm{Proj}_{\cD}\left(\hat\delta^{(k)} + \eta\lambda\frac{1}{M} \sum_{k \in [M]} z^{(k)}\right)$}
Return $\hat\delta^{(\infty)} = \hat\delta^{\rm{MCMC}}$.
\end{algorithm}

\section{Additional Numerical Experiments}\label{sec:app_additional_experiments}
The code is provided in the supplementary material.
\subsection{Numerical Discussions on  \texorpdfstring{\cref{assumption:contractance}}{Lg}}\label{ref:assumption_comments}

\begin{figure}
    \centering
    \includegraphics[width=\linewidth]{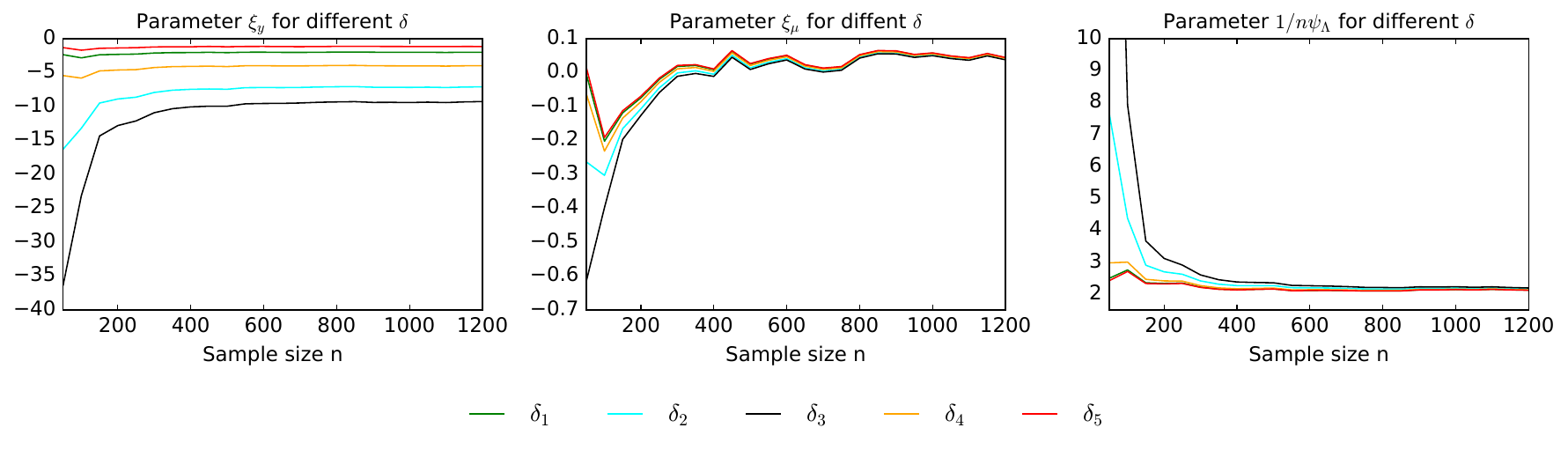}
    \caption{Convergence of variational parameters $(\xi_y, \xi_\mu, \psi_\Lambda)$ with respect to the sample size $n$ in dimension $d=1$ for the GW model. We take $5$ different values of $\delta$ randomly (5 different colors).}
    \label{fig:contractance_Assumption}
\end{figure}

We evaluate the convergence of the variational parameters as the sample size $n$ increases. Specifically, we generate synthetic data with dimension $d=1$ under the GW model, and for values of $n \in [50, 1200]$, we compute the variational parameters $(\xi_y, \xi_\mu, \psi_\Lambda)$ for different random values of $\delta$, using the corresponding fixed-point computation. \cref{fig:contractance_Assumption} illustrates that the variational distribution derived from the fixed-point equation converges to the variational distribution obtained from the asymptotic fixed-point operator.

\subsection{Number of Fixed-point Iterations}
\begin{figure}
    \centering
    \includegraphics[width=0.8\linewidth]{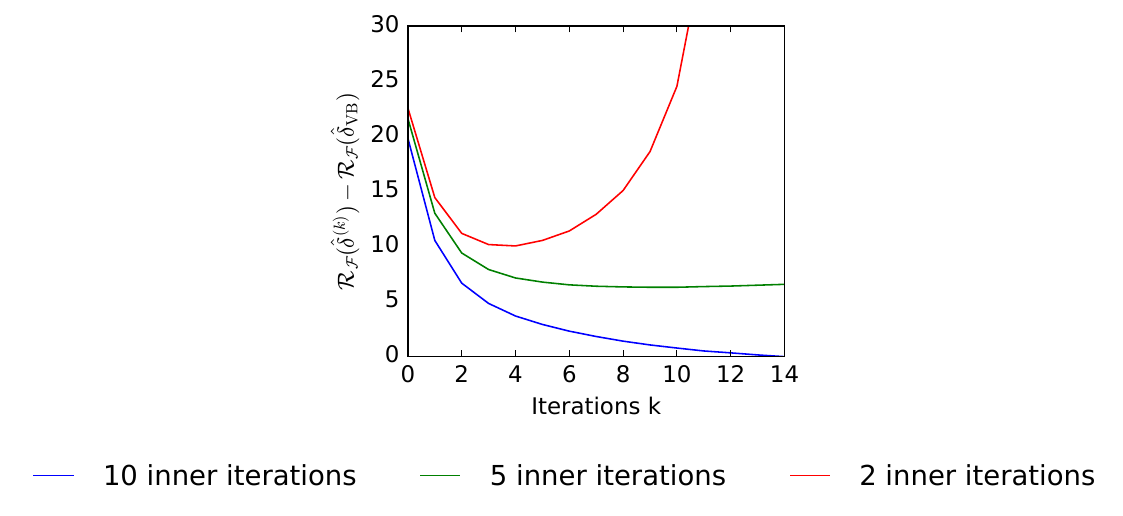}
    \caption{Convergence of the difference $\obj(\deltaVB) - \obj(\delta^{(k)})$ with respect to the $k^{\rm{th}}$ iteration of the algorithm on the GW model. We repeat the experiment for a different amount of inner iterations. We set $d=50$.}
    \label{fig:inner_iterations}
\end{figure}

We evaluate the difference in the value function, $\obj(\deltaVB) - \obj(\delta^{(k)})$, of our algorithm when applied to the GW model on a synthetic dataset. We set $d=50$ and fix $n=200$, and examine the impact of the number of inner iterations performed on the fixed-point computation. \cref{fig:inner_iterations} demonstrates that insufficient inner iterations result in non-convergence of the value function $\obj(\hat\delta^{(k)})$, as the supremum in the objective function is not reached.

\subsection{Computational Complexities}
By employing the mean-field assumption, our algorithm \algVB~ simplifies the optimization problem from a measure-based setting (\cref{eq:final_min_max}) to a finite-dimensional parametric optimization problem.
The primary computational burden lies in the inversion of $(d, d)$ matrices for both the stationary and autoregressive Gaussian-Wishart models, resulting in a computational complexity of $\mathcal{O}(d^3)$. In contrast, the Gaussian Process (GP) model faces scalability challenges, as it requires the inversion of $(nd, nd)$ matrices, leading to cubic complexity with respect to both dimension $d$ dataset size $n$. Existing works such as those proposed by \citet{quinonero2005unifying} aim at reducing this inversion complexity; we leave this challenging task as future work.

\vfill

\end{document}